\documentclass[amsmath,amssymb,12pt,superscriptaddress,reprint, preprintnumbers, notitlepage,aps,onecolumn,nofootinbib]{revtex4-2}
\pdfoutput=1 
\usepackage[utf8]{inputenc}
\usepackage[english]{babel}
\usepackage{amsmath}
\usepackage{graphicx}
\usepackage{dcolumn}
\usepackage{pbox}
\usepackage{amssymb}
\usepackage{epsfig}
\usepackage{slashed}
\usepackage{amssymb}
\usepackage{ mathrsfs }
\usepackage{color}
\usepackage{url}
\usepackage{multirow}
\def\simlt{\stackrel{<}{{}_\sim}}
\def\simgt{\stackrel{>}{{}_\sim}}
\usepackage{lineno}
\usepackage[table]{xcolor}
\usepackage{xcolor,pifont}
\usepackage{comment}
\usepackage{mathtools}
\usepackage{float}
\usepackage{multirow}
\usepackage{comment}
\usepackage{caption}
\usepackage[hidelinks]{hyperref}

\newcommand{\g}{G_{\hbox{\tiny{NC}}}}
\newcommand{\gd}{\hat{G}_{\hbox{\tiny{NC}}}}
\newcommand{\G}{\mathfrak{g}_{\hbox{\tiny{NC}}}}

\DeclarePairedDelimiter\ket{\lvert}{\rangle}
\DeclarePairedDelimiter\bra{\langle}{\rvert}
\begin{document}
\title{\bf Quantum simulation of quantum mechanical system with spatial noncommutativity}
\author{S. Hasibul Hassan Chowdhury}
\email[E-mail: ]{shhchowdhury@bracu.ac.bd}
\affiliation{Department of Mathematics and Natural Sciences, BRAC University, Dhaka 1212, Bangladesh}
\author{Talal Ahmed Chowdhury}
\email[E-mail: ]{talal@du.ac.bd}
\affiliation{Department of Physics, University of Dhaka, Dhaka 1000, Bangladesh}
\affiliation{Department of Physics and Astronomy, University of Kansas, Lawrence, Kansas 66045, USA}
\affiliation{The Abdus Salam International Centre for Theoretical Physics, Strada Costiera 11, I-34014, Trieste, Italy}
\author{Salah Nasri}
\email[E-mail: ]{snasri@uaeu.ac.ae}
\affiliation{Department of Physics, UAE University, P.O. Box 17551, Al-Ain, United Arab Emirates}
\affiliation{The Abdus Salam International Centre for Theoretical Physics, Strada Costiera 11, I-34014, Trieste, Italy}
\author{Omar Ibna Nazim}
\email[E-mail: ]{omaribnanazim@gmail.com}
\affiliation{Department of Physics, University of Dhaka, Dhaka 1000, Bangladesh}
\author{Shaikh Saad}
\email[E-mail: ]{shaikh.saad@unibas.ch}
\affiliation{Department of Physics, University of Basel, Klingelbergstrasse\ 82, CH-4056 Basel, Switzerland}

\begin{abstract}
    Quantum simulation has become a promising avenue of research that allows one to simulate and gain insight into the models of High Energy Physics whose experimental realizations are either complicated or inaccessible with current technology. We demonstrate the quantum simulation of such a model, a quantum mechanical system with spatial noncommutativity, which is inspired by the works in Noncommutative Geometry and Noncommutative Field theory for a universal quantum computer. We use the novel group theoretical formalism to map the Hamiltonian of such a noncommutative quantum system into the ordinary quantum mechanical Hamiltonian and then carry out the quantum simulation using the Trotter-Suzuki product formula. Furthermore, we distinguish the impact of the noncommutativity parameter on the quantum simulation, especially on the Trotter error, and point out how its sizable value affects the simulation.
\end{abstract}
\maketitle

\section{Introduction}

Quantum simulation, the idea put forward in the pioneering works \cite{benioff1980computer, manin, feynman}, describes the simulation of a seemingly complicated quantum system with another well-controlled quantum system. It has become an active field of research \cite{cirac2012goals, Georgescu:2013oza, Daley:2022eja} and drawing much attention from the high energy physics community \cite{Bauer:2022hpo, Humble:2022vtm, Catterall:2022wjq, Alam:2022crs, Spentzouris:2020ilm}. The quantum simulator can be classified broadly into two classes: analog and digital simulators. The analog simulator involves a relatively isolated and controlled quantum system in the laboratory setup whose constituent degrees of freedom resemble the degrees of freedom of the underlying quantum system one simulates. On the other hand, the digital simulator is a digital quantum computer that operates on an array of two-level quantum systems known as qubits, following a set of universal and elementary operations. Besides, the attributes of both analog and digital simulators can be integrated for a more efficient and adaptable quantum simulation of a complex quantum system. Although the quantum simulation was first prescribed for the many-body Hamiltonian in \cite{Lloyd} and subsequently improved in \cite{weisner, Lloyd-Abrams, somaroo, zalka, farhi, ortiz, somma-ortiz, somma-ortiz-knill, berry, childs}, its importance was quickly realized for areas related to the High Energy Physics (HEP). For example, the possibility of simulating physics of many degrees of freedom, especially the lattice gauge theories using cold atoms and trapped ions, has been shown in \cite{cirac-zollar, james, jaksch, deutsch-brennen, lewenstein, buluta, casanova, lanyon, bloch2012quantum, blatt, Wiese:2013uua, zohar, carmen, Monroe, florian, monika}. In addition, the quantum algorithm to simulate the quantum field theory with a universal quantum computer has been presented in \cite{preskill-0, preskill-1, preskill-2, Jordan:2017lea, preskill-3, HamedMoosavian:2017koz, harnik-1, harnik-2, Klco-1, Hackett:2018cel, Yeter-Aydeniz:2018mix, Kreshchuk:2020kcz, Kreshchuk:2020dla, Haase:2020kaj, Liu:2020eoa, Stetina:2020abi, Klco:2021lap}. Moreover, the implementation of the lattice gauge theories in the quantum computer has been carried out in \cite{byrnes, Muschik:2016tws, Lamm:2019bik, Alexandru:2019nsa, Harmalkar:2020mpd, Gustafson:2020yfe, Carena:2021ltu, Carena:2022kpg, Ciavarella:2022zhe, Halimeh:2022mct}. Apart from that, the application of quantum simulation and quantum computing have been found in the determination of the Parton distribution functions \cite{Lamm:2019uyc, Li:2021kcs, Perez-Salinas:2020nem, Bepari:2020xqi, Echevarria:2020wct,Kreshchuk:2020aiq,Kirby:2021ajp}, and the simulation of the Parton shower that captures the quantum effects \cite{Bauer:2019qxa, Bauer:2021gup, Gustafson:2022xwt}. Also, the quantum simulation has been used to improve the jet clustering algorithm \cite{Wei:2019rqy, Pires:2020urc, Pires:2021fka, deLejarza:2022bwc, Delgado:2022snu}, to simulate full medium induced Parton shower and to probe jet quenching \cite{Barata:2021yri, Barata:2022wim, Yao:2022eqm}, and the dynamics of the heavy quark or jet in strongly coupled quark-gluon plasma in heavy ion collision \cite{DeJong:2020riy, Cohen:2021imf}. Furthermore, the usage of the quantum algorithm can facilitate more robust data analysis in HEP \cite{Delgado:2022tpc}, reconstruct the particle tracks more efficiently \cite{Tuysuz:2020ocw}, extract physical observables from scattering processes \cite{Martens:2017cvj, Kharzeev:2021nzh}, and it can be relevant for future colliders \cite{Gray:2022fou}. A Quantum algorithm has also been developed to evaluate the Feynman loop integral \cite{Ramirez-Uribe:2021ubp}, which is essential for precision HEP calculations. The quantum simulation of the phase transition in gauge theories has been demonstrated in \cite{Czajka:2021yll, Cohen:2021imf, Davoudi:2022uzo} as well. As we have entered the era of noisy intermediate-scale Quantum (NISQ) devices \cite{preskill_NISQ, Bharti:2021zez}, by exploring the quantum simulation of the lower-dimensional field theories, for example, the Schwinger model \cite{Schwinger:1962tp, Lowenstein:1971fc, Coleman:1975pw, Coleman:1976uz} in series of works \cite{Hauke:2013jga, Kuhn:2014rha, Ercolessi:2017jbi, Chakraborty:2020uhf, Shaw:2020udc, Nguyen:2021hyk, Honda:2021ovk, Honda:2021aum, deJong:2021wsd, Thompson:2021eze, Xie:2022jgj}, and Gross-Neveu model \cite{gross-neveu} in \cite{Asaduzzaman:2022bpi}, one can interpolate how to address the simulation of the full $3+1$ dimensional quantum field theories in the NISQ devices and future quantum computers. Likewise, the quantum simulation of the $O(3)$ Sigma model has been prescribed in \cite{Schutzhold:2004xq, Alexandru:2019ozf, Singh:2019uwd, Buser:2020uzs}. Additionally, the procedure to simulate the non-perturbative processes like Schwinger pair production in the strong electric field in a quantum computer has been presented in \cite{Xu:2021tey}. Therefore, one can see the rapid development of algorithms and devices associated with quantum simulation for HEP-related research areas.  

Another interesting feature of quantum simulation is its capability to simulate and provide insight into the novel quantum many-body system and the exotic phases of matter whose experimental realizations are either difficult or inaccessible with current technology. One such example is the time crystal \cite{Wilczek:2012jt, else-floquet, khemani}, which is the phase of the system arising from the spontaneous breaking of time-translation symmetry, and it only exists when the system is in a complete non-equilibrium state that seemingly contradicting our usual notion of the stability of the phase of a many-body system. Such time crystal phase has been observed in quantum simulator \cite{zhang2017observation, floquet-cold-atom, randall-solid-state-spin, mi2022time, Ying, xu2021realizing}. Besides, many-body dynamics, especially localization phenomena and the phase transitions, are simulated in \cite{choi, mazurenko, bernien, zhang-pagano}. In addition, the demonstration of the quantum phase transition has been presented in \cite{scholl, ebadi, bartlett, satzinger, semeghini, mei}, and the exotic topological phases are probed in \cite{dumitrescu, zhang2022digital, Kiczynski:2022jmm, Ashhab-njp} by using quantum simulators. Hence, quantum simulation can offer ways to enhance our physical understanding of quantum systems which are 
beyond the reach of the contemporary experimental setup. 

In this work, we consider one other example of such a system, the quantum mechanical system with spatial noncommutativity, generally known as noncommutative quantum mechanics (NCQM) \cite{Nair:2000ii, Gamboa:2000yq}, the prototype model that is inspired by the works on the noncommutative geometry \cite{connes1995noncommutative} and noncommutative field theory (for reviews, see Refs.~\cite{Douglas:2001ba, Szabo:2001kg} and references therein). In the NCQM, the canonical commutation relations between the respective position and momenta operators accommodate spatial as well as momentum noncommutativity. The position noncommutativity leads to the fuzziness of points in the spatial context, while the momentum noncommutativity signals the presence of a background magnetic field (see \cite{Delduc:2007av} for a detailed account). Although probing the signatures of spatial noncommutativity requires access to a very high energy scale \cite{Carroll:2001ws, Szabo:2009tn}, which is currently not possible for any terrestrial experiments, there are suggestions to test such noncommutativity in an artificial analogue system using cold Rydberg atoms \cite{Zhang:2004yu}. Besides, the indirect effects of quantized spacetime, such as nonlocal interactions, can be simulated in an array of atomic ensembles within an optical cavity \cite{Periwal:2021eur}, and the low-energy imprints of some models of quantum gravity can be probed in well-controlled matter-wave interferometry, and optomechanical systems \cite{Carney:2018ofe}. On the contrary, the noncommutative geometrical description can be realized in condensed matter system, for example, in the case of Landau levels \cite{Jackiw:2001dj, Jackiw:2002wd}, integer quantum hall effect \cite{bellissard}, fractional quantum hall effect \cite{Haldane:2011ia} and topological insulators \cite{prodan, neupert, schulz-baldes}. Therefore, we are interested in simulating the dynamics of a two-dimensional noncommutative quantum mechanical system in a quantum computer. While implementing the NCQM in the quantum simulator, we have used the group theoretical formalism developed in \cite{Chowdhury:2012ik, Chowdhury:2013cca, Chowdhury:2015gsk, Chowdhury:2015hnp}, which is based on constructing the families of unitary irreducible representations of the kinematical symmetry group associated with the NCQM system. We also want to emphasize that, although, in the usual picture of NCQM where the noncommutativity between the fundamental positions and momenta are considered, one can go beyond and address an effective description of a quantum system whose dynamics can be captured by a suitably constructed set of noncommutative dynamical variables. Hence, as a starter, we have carried out the quantum simulation of a two-dimensional noncommutative quantum system so that we can identify how it can be used to probe the additional noncommutativity parameter(s) of a quantum mechanical system.

This article is organized as follows. In section \ref{ncqmsec}, we present the group theoretical construction of the two-dimensional noncommutative quantum mechanical system. The details of the quantum algorithm for the quantum simulation are given in section \ref{quantumalgo}. We discuss the results of the quantum simulation and its limitations in section \ref{discussion} and conclude in section \ref{conclusion}.

\section{Noncommutative quantum mechanical system}\label{ncqmsec} 
\subsection{Group theoretical formalism for the NCQM}\label{grouptheoretical}
The quantum phase space associated with the commutative two-dimensional quantum mechanical system is described by the position and momenta operators, which are unbounded Hermitian operators acting on the associated Hilbert space. For a two-dimensional system, the classical phase space consists of the position coordinates $x$ and $y$ and the respective momentum coordinates $p_x$ and $p_y$. Now the corresponding quantum phase space coordinates comprise of the Hermitian operators $\hat{x}$, $\hat{y}$, $\hat{p}_{x}$, $\hat{p}_{y}$ defined on $L^{2}(\mathbb{R}^{2},dx\;dy)$ and satisfying the following commutation relations that correspond to the Weyl-Heisenberg group,
\begin{align}
&[\hat{x},\hat{p}_{x}]=[\hat{y},\hat{p}_{y}]=i\hbar\mathbb{I},\nonumber\\
&[\hat{x},\hat{y}]=[\hat{p}_{x},\hat{p}_{y}]=0, \label{Heisenberg-algebra}
\end{align}
where $\mathbb{I}$ is the identity operator on $L^{2}(\mathbb{R}^{2},dx\;dy)$.

Now consider the two-dimensional noncommutative plane immersed in a constant magnetic field $B$ where the spatial noncommutativity is parameterized by the constant $\theta$. In group theoretical formulation, one first starts with a two-dimensional configuration space with coordinates $(q_{1},q_{2})$. One subsequently forms the associated 4-dimensional phase space, the coordinates of which can be read off from the ordered 4-tuple $(q_{1},q_{2},p_{1},p_{2})$ where $p_{i}$ are the conjugate momenta associated with $q_{i}$. The underlying physical system's time evolution is described by a curve parameterized by time $t$ in the underlying phase space. The evolving physical system follows a trajectory in phase space subjected to some initial condition. Now the underlying system is expected to possess translational symmetry. The translational symmetry group $T_{(Q_{1},Q_{2},P_{1},P_{2})}$ acts on the phase space point $(q_{1},q_{2},p_{1},p_{2})$ describing the system at a particular instant in the following way:
\begin{equation}\label{group-action}
T_{(Q_{1},Q_{2},P_{1},P_{2})}(q_{1},q_{2},p_{1},p_{2})=(q_{1}+Q_{1},q_{2}+Q_{2},p_{1}+P_{1},p_{2}+P_{2}).
\end{equation}
This set of transformations form   a 4 dimensional abelian Lie group, with the  group multiplication  given as follows:
\begin{equation}\label{grp-mult-translation-symmetry}
T_{(Q_{1},Q_{2},P_{1},P_{2})}T_{(Q^{\prime}_{1},Q^{\prime}_{2},P^{\prime}_{1},P^{\prime}_{2})}=T_{(Q_{1}+Q^{\prime}_{1},Q_{2}+Q^{\prime}_{2},P_{1}+P^{\prime}_{1},P_{2}+P^{\prime}_{2})}.
\end{equation}
All of its irreducible unitary representations are one-dimensional representations ($\mathbb{C}$ is the representation space), and the group generators are represented by constant multiples of identity operator on $\mathbb{C}$. 

At this stage, using the formalism of \cite{bargmann1954unitary}, we centrally extend the above-mentioned  4-dimensional translational symmetry group by $\mathbb{R}^3$ so that the center of the resulting centrally extended nonabelian group is $\mathbb{R}^3$. We denote this 7-dimensional real nilpotent Lie group by $\g$ and its Lie algebra by $\G$. It has been established as the kinematical symmetry group of 2-dimensional noncommutative quantum mechanics in \cite{Chowdhury:2013cca,Chowdhury:2015gsk}. If one denotes a generic element of $\g$ by $(\Theta,\Phi,\Psi,Q_{1},Q_{2},P_{1},P_{2})$, then the corresponding group multiplication in $\g$ reads (\cite{Chowdhury:2013cca})
\begin{eqnarray}\label{group-law-G_NC}
\lefteqn{(\Theta,\Phi,\Psi,Q_{1},Q_{2},P_{1},P_{2})(\Theta^{\prime},\Phi^{\prime},\Psi^{\prime},Q^{\prime}_{1},Q^{\prime}_{2},P^{\prime}_{1},P^{\prime}_{2})}\nonumber\\
&&=(\Theta+\Theta^{\prime}+\frac{\alpha}{2}(Q_{1}P^{\prime}_{1}+Q_{2}P^{\prime}_{2}-P_{1}Q^{\prime}_{1}-P_{2}Q^{\prime}_{2}),\Phi+\Phi^{\prime}+\frac{\beta}{2}(P_{1}P^{\prime}_{2}-P_{2}P^{\prime}_{1}),\\ &&\hspace{.2in}\Psi+\Psi^{\prime}+\frac{\gamma}{2}(Q_{1}Q^{\prime}_{2}-Q_{2}Q^{\prime}_{1}),Q_{1}+Q^{\prime}_{1},Q_{2}+Q^{\prime}_{2},P_{1}+P^{\prime}_{1},P_{2}+P^{\prime}_{2}).\nonumber
\end{eqnarray}
Here $(\Theta,\Phi,\Psi)\in\mathbb{R}^{3}$ which is the centre of $\g$, and $\alpha$, $\beta$ and $\gamma$ are some fixed constants carrying dimensions of $(\hbox{momentum}\times\hbox{position})^{-1}$, $(\hbox{momentum})^{-2}$ and $(\hbox{position})^{-2}$, respectively.

The dual Lie algebra $\G^{*}$ is also 7-dimensional. There is a natural action of $\g$ on  $\G^{*}$ known as coadjoint action. The orbits of the coadjoint action are called coadjoint orbits. The 7-dimensional vector space $\G^{*}$ is found to be foliated by coadjoint orbits of 3 different dimensions as found in \cite{Chowdhury:2013cca}: 4 dimensional, 2 dimensional, and 0 dimensional. By a theorem due to Kirillov \cite{kirillov}, the coadjoint orbits of $\g$ are in one-to-one correspondence with the unitary dual $\gd$, i.e., the set of equivalence classes of unitary irreducible representations of $\g$. A generic element of $\gd$ is given by an ordered triple $(\hbar,\theta,B)$ where $\theta$ and $B$ are the spatial noncommutativity parameter and the constant magnetic field, respectively. The noncentral generators corresponding to the group parameters $Q_{1}$, $Q_{2}$, $P_{1}$ and $P_{2}$ are represented by the 2-parameter $(r,s)$ family of self-adjoint operators $\hat{Q}^{s}_{1}$, $\hat{Q}^{s}_{2}$, $\hat{\Pi}^{r,s}_{1}$, $\hat{\Pi}^{r,s}_{2}$, respectively acting on $L^{2}(\mathbb{R}^{2},dx\;dy)$ subjected to the following set of commutation relations:
\begin{equation}\label{commutation-G_NC}
[\hat{Q}^{s}_{i},\hat{\Pi}^{r,s}_{j}]=i\hbar\delta_{ij}\hat{\mathbb{I}},\;[\hat{Q}^{s}_{1},\hat{Q}^{s}_{2}]=i\theta\hat{\mathbb{I}},\;[\hat{\Pi}^{r,s}_{1},\hat{\Pi}^{r,s}_{2}]=i\hbar B\hat{\mathbb{I}}.
\end{equation}
Here $\hat{\mathbb{I}}$ is the identity operator on $L^{2}(\mathbb{R}^{2},dx\;dy)$. The quantum phase space comprising the 2-parameter family of self-adjoint operators $\{\hat{Q}^{s}_{1},\hat{Q}^{s}_{2},\hat{\Pi}^{r,s}_{1},\hat{\Pi}^{r,s}_{2}\}$ subject to the commutation relations Eq. (\ref{commutation-G_NC}) physically represents the noncommutative quantum mechanical plane immersed in a constant magnetic field. Therefore, we can see that the parameters characterizing the spatial and momentum noncommutativity, $\theta$ and $B$, respectively, along with the usual quantum mechanical position-momentum noncommutativity given by $\hbar$, arise from the central extension of the associated kinematical symmetry group of the system. 

The map that expresses the noncommutative operators $\hat{Q}^{s}_{1}$, $\hat{Q}^{s}_{2}$, $\hat{\Pi}^{r,s}_{1}$, $\hat{\Pi}^{r,s}_{2}$ in terms of the quantum mechanical operators $\hat{x},\,\hat{y},\, \hat{p}_{x},\,\hat{p}_{y}$ is given by,
\begin{equation}\label{NCQMtoQM}
\begin{aligned}
 \hat{Q}^{s}_{1}&=\hat{x}-s\dfrac{\theta}{\hbar}\hat{p}_{y},\,\,\, \hat{Q}^{s}_{2}=\hat{y}+(1-s)\dfrac{\theta}{\hbar}\hat{p}_{x},\\
 \hat{\Pi}^{r,s}_{1}&=\dfrac{B\hbar(1-r)}{\hbar-B\theta r}\hat{y}+\dfrac{[B\theta(r+s-rs)-\hbar]}{B\theta r-\hbar}\hat{p}_{x},\\
 \hat{\Pi}^{r,s}_{2}&=-rB\hat{x}+\left[1+r(s-1)\dfrac{B\theta}{\hbar}\right]\hat{p}_{y}.
\end{aligned}
\end{equation}
where the 2-parameters $(r,s)$ represent the family of unitary equivalent irreducible representations of $\g$ associated with the fixed ordered triple $(\hbar,\theta,B)$. Now  two representations labeled by $(r,s)$ and $(r^{\prime},s^{\prime})$ are intertwined by a unitary operator $U$ on the given Hilbert space $L^{2}(\mathbb{R}^{2},dx\;dy)$ where the group generators also transform, using the same unitary operator $U$, as follows
\begin{equation}
\begin{aligned}
&\hat{Q}_{1}^{s^{\prime}}=U\hat{Q}_{1}^{s}U^{-1},\,\,\,\hat{Q}_{2}^{s^{\prime}}=U\hat{Q}_{2}^{s}U^{-1},\\
&\hat{\Pi}_{1}^{r^{\prime},s^{\prime}}=U\hat{\Pi}^{r,s}_{1}U^{-1},\,\,\hat{\Pi}_{2}^{r^{\prime},s^{\prime}}=U\hat{\Pi}^{r,s}_{2}U^{-1},
\end{aligned}
\end{equation}
leading to the following unitary transformation of the underlying Hamiltonian: 
\begin{equation}\label{eq:Hamiltonian-unitary-transformation}
\hat{H}^{r^{\prime},s^{\prime}}=U\hat{H}^{r,s}U^{-1}.
\end{equation}
One can consider the parameters $(r,s)$ as the gauge parameters as they do not affect the measurable quantities, for example, the energy spectra of the noncommutative Hamiltonian as shown in \cite{Chowdhury:2020oue}.

\textbf{A possible generalization:} One can encounter a quantum system that is described not by the usual position and momentum Hermitian operators but by a set of $2n$ Hermitian operators $\hat{A}_{i}$ and $\hat{B}_{i}$ associated with dynamical variables $a_{i}$ and $b_{i}$ with $i=1,2,..,n$, respectively which effectively characterize the system. The set of Hermitian operators $(\hat{A}_{i},\, \hat{B}_{i})$ follow the following set of commutation relations: $[\hat{A}_{i},\hat{A}_{j}]=i\theta_{ij}$, $[\hat{B}_{i},\hat{B}_{j}]=i\gamma_{ij}$ and $[\hat{A}_{i},B_{j}]=i \alpha_{ij}$ where $\theta_{ij}$ and $\gamma_{ij}$ are the matrix entries of some real skew-symmetric $n\times n$ matrices $\theta$ and $\gamma$, respectively. Also, $\alpha_{ij}$ are the entries of the real diagonal matrix $\alpha$. The constant real parameters $\alpha_{ij}$, $\theta_{ij}$ and $\gamma_{ij}$ are associated with the noncommutativity of the operators $\hat{A}_{i}$ and $\hat{B}_{j}$ for all $i,j=1,2,..n$. Following the group theoretical formalism, first, we identify the configuration space $\mathcal{M}$ as a smooth manifold that is locally described by the dynamical variables $a_{i}$. One then enumerates the respective conjugate variables $b_{i}$ so that the $2n$ local coordinates $(a_{i},b_{i})$ describe the underlying cotangent bundle $T^{*}\mathcal{M}=:\mathcal{N}$. Here, $\mathcal{N}$ is a real $2n$-dimensional Poisson manifold and is naturally endowed with a symplectic structure. Besides, one can go on to define a Hamiltonian function $H:\mathcal{N}\rightarrow\mathbb{R}$ that can be considered the generator of one-parameter transformation on points of $\mathcal{N}$. At this point, one tries to identify the kinematical symmetry group $G_{\mathcal{N}}$ associated with the manifold $\mathcal{N}$ and centrally extend it to $G^{\mathrm{ext}}_{\mathcal{N}}$ whose center is $\mathbb{R}^{N}$ where $N$ is the number of noncommutativity parameters associated with the initial quantum system. In this way, one constructs the kinematical symmetry group $G^{\mathrm{ext}}_{\mathcal{N}}$ of the underlying quantum system. Eventually, one can construct the equivalence classes of unitary irreducible representations of $G^{\mathrm{ext}}_{\mathcal{N}}$ that are associated with the $(2n\times 2n)$ real matrix $\Omega$ of noncommutativity written in the block form as $\Omega=\begin{bmatrix}\theta&\alpha\\-\alpha&\gamma\end{bmatrix}$. Moreover, the generators associated with the noncentral group parameters $a_{i}$ and $b_{i}$ are nothing but the Hermitian operators $\hat{A}_{i}$ and $\hat{B}_{i}$ acting on some suitable Hilbert Space that follow the above-mentioned commutation relations. Therefore, the noncommutative two-dimensional quantum mechanical system can be an example of the generalized group theoretical construction.

\subsection{Noncommutative two-dimensional oscillator}\label{ncoscillator}
We consider simulating the noncommutative two-dimensional harmonic oscillator \cite{Hatzinikitas:2001pm, Smailagic:2001qe, Muthukumar:2002cn}, one of the representatives of the noncommutative two-dimensional quantum mechanical system. The Hamiltonian that describing a particle of mass $m$ subjected to a constant magnetic field $B$ along the perpendicular direction of the plane and an anisotropic harmonic potential, $V(\hat{Q}_{1},\hat{Q}_{2})=\frac{1}{2}m\big[\omega_1^2\hat{Q}_{1}^2+\omega_2^2\hat{Q}_{2}^2\big]$, is given as
\begin{equation}\label{eq:NC-Hamiltonian}
\hat{H}^{NC}=\frac{1}{2m}\big[\hat{\Pi}_{1}^2+\hat{\Pi}_{2}^2\big]+\frac{1}{2}m\big[\omega_1^2\hat{Q}_{1}^2+\omega_2^2\hat{Q}_{2}^2\big].
\end{equation}
Using Eq. (\ref{NCQMtoQM}), we can obtain the following form of the noncommutative Hamiltonian expressed in terms of the quantum mechanical operators;
\begin{equation}\label{commhamiltonian}
\begin{aligned}
\hat{H}=\frac{1}{2M_1}\hat{p}_{x}^2+\frac{1}{2M_2}\hat{p}_{y}^2+\frac{1}{2}M_1\Omega_1^2\hat{x}^2+\frac{1}{2}M_2\Omega_2^2\hat{y}^2-l_1\hat{x}\hat{p}_{y}+l_2\hat{y}\hat{p}_{x}.
\end{aligned}
\end{equation}
where, 
\begin{align}
    M_{1}& = \left[\frac{(\hbar-(r+s-r s)\theta B)^2}{m(\hbar-r\theta B)^{2}}+\frac{m(1-s)^2\theta^2\omega_{2}^2}{\hbar^2}\right]^{-1},\,\,\,
    M_{2}= \left[\frac{1}{m}\left(1-\frac{(1-s)r\theta B}{\hbar}\right)^{2}+\frac{m s^2\theta^2\omega_{1}^{2}}{\hbar^{2}}\right]^{-1}\nonumber\\
    \Omega_{1}^{2}&=\frac{m}{M_{1}}\left(\omega_{1}^{2}+\frac{B^{2}r^{2}}{m^{2}}\right),\,\,\,
    \Omega_{2}^{2}=\frac{m}{M_{2}}\left(\omega_{2}^{2}+\frac{B^{2}(1-r)^{2}\hbar^{2}}{m^{2}(\hbar-r\theta B)^{2}}\right),\nonumber\\
    l_{1}&=\frac{m s \theta \omega_{1}^{2}}{\hbar}+\frac{B r}{m}\left(1-\frac{(1-s)r\theta B}{\hbar}\right),\,\,\, l_{2}= \frac{m (1-s) \theta \omega_{2}^{2}}{\hbar}+\frac{\hbar B(1-r)(\hbar+\theta B(r s - r -s))}{m(\hbar-r\theta B)^{2}}.
    \label{parameters}
\end{align}
Here, we can see that the initial parameters of the NC Hamiltonian Eq. (\ref{eq:NC-Hamiltonian}), $\{m,\,\omega_{i},\,\hbar,\,\theta,\, B,\,r,s\}$ enter into the parameters, $\left(M_{i},\Omega_{i},l_{i}\right)$ given in Eq. (\ref{parameters}) of Eq. (\ref{commhamiltonian}).
Besides, in the case of zero magnetic field, $B=0$, the family of unitary equivalent irreducible representations of $\g$ is only parameterized by $s$, and are associated with the fixed ordered doublet $(\hbar,\theta)$. Moreover, in this case, the noncommutative momentum operators, $\hat{\Pi}_{1,2}$ become just the usual quantum mechanical momentum operators, $\hat{p}_{x,y}$ and $r$ parameter drops out in the relations Eq. (\ref{parameters}). In the following sections, we address the quantum simulation of the noncommutative isotropic two-dimensional harmonic oscillator in the zero magnetic field for its simplicity and leave the case of the anisotropic harmonic oscillator in a non-zero magnetic field for separate work.

\section{Quantum algorithm}\label{quantumalgo}
To study the evolution of a quantum system first, we choose a basis of the Hilbert Space, which is generally associated with a complete set of mutually commuting observables. Afterward, we find a suitable expression of the time evolution operator in that basis and evolve an initial quantum state by applying the time evolution operator to obtain the final state. When we measure one of the observables, the final state will collapse onto an eigenstate of that observable with a specific probability. If the time evolution is simulated multiple times, we obtain the probability distribution of the final state. The qubit is the fundamental computing element of the quantum computer, which is a  two-state quantum system. So a system of $n$ qubits will represent the Hilbert Space of $2^n$ dimensions and constitute the computational basis on which the quantum computer operates. For performing the quantum simulation of a quantum system, first, we must establish a mapping between the eigenbasis for a relevant observable and the computational basis corresponding to $n$ qubits where $n$ is chosen in such a way that we can establish the one-to-one mapping between the eigenbasis and computational basis. The next step is to find the representation of the time evolution operator on a computational basis. The time evolution operator will be decomposed into unitary operators known as the quantum gates, which are applied to the qubits. At this point, we are ready to execute the simulation in a quantum computer. First, we prepare an initial quantum state, and then by applying appropriate quantum gates to it, we evolve that state in time to obtain the final state. Subsequently, we implement the measurement of the final state and note the resultant computational basis state. In this way, by repeating the simulation many times and making measurements of the final states, we obtain the probability distributions of the computational basis states. Thereupon, we retrieve their probability distributions and interpret the results by using the inverse mapping from the computational basis to the relevant eigenbasis of the chosen observable. For the quantum simulation of the noncommutative two-dimensional isotropic harmonic oscillator, we extend the algorithm of \cite{somma2015quantum} in the case of two-dimensional quantum mechanical systems. In the following sections, we describe the steps in detail.

\subsection{Mapping to computational basis states}\label{compbasis} 
In our simulation, we discretize the two-dimensional plane into a mesh by dividing each dimension of the plane into $N$ divisions being labeled from $-N/2$ to $N/2-1$ with a unit interval, and each discrete point of this $N\times N$ mesh is considered to represent a position eigenstate. Besides, we choose $N$ to be equal to $2^p$ so that  $p=\mathrm{log}_{2}N$ number of qubits are needed to obtain the same number of computational states along one direction, for example, $x$ axis, and therefore the total number of qubits to represent $|x, y\rangle$ position eigenstate will be $2p$. The computational basis state corresponding to the position eigenstate $|x,y\rangle$ is obtained by associating $x=i+N/2$ and $y=j+N/2$ to $\mathrm{bin}(i+N/2)$ and $\mathrm{bin}(j+N/2)$, respectively, for $i,\,j=-N/2,\,-N/2+1,\,...,\,N/2-2,\,N/2-1$ where $\mathrm{bin}(a)$ denotes binary of the decimal number $a$ in $p$ bits. Therefore, the position eigenstate $|x,y\rangle$ is expressed as $\ket{\mathrm{bin}(i+N/2)}\otimes \ket{\mathrm{bin}(j+N/2)}\equiv\ket{\mathrm{bin}(i+N/2)\mathrm{bin}(j+N/2)}$. For example, if we divide $x-y$ plane into $8\times 8$ mesh with $N=8$ so that $p=\mathrm{log}_{2}8=3$ and accordingly the $|-4,-4\rangle$ position state is mapped to computational basis $\ket{\mathrm{bin}(0) \mathrm{bin}(0)}\equiv\ket{000000}$, $|-4,-3\rangle$ to $\ket{\mathrm{bin}(0) \mathrm{bin}(1)}\equiv\ket{000001}$, $|0,0\rangle$ to $\ket{\mathrm{bin}(4) \mathrm{bin}(4)}\equiv\ket{100100}$ and so on.

\subsection{Operator representations in the computational basis}\label{operatorcompbasis}
In the discrete setup, the position operators are diagonal matrices, and corresponding momentum operators, which will also be finite-dimensional matrices, can be obtained using the centered quantum Fourier transform \cite{nielsen_chuang_2010}. The quantum mechanical position operators $\hat{x}$ and $\hat{y}$, now represented by $N\times N$ diagonal matrices and denoted as $X$ and $Y$, respectively, are given as,
\begin{equation}
X=Y = \sqrt{\frac{2\pi}{N}}
\begin{pmatrix}
-N/2 & 0 & .& . &  .& 0\\
0 &-N/2+1 & .& . & . & 0 \\
.&.&.& &&.\\
.&.&& .&&.\\
.&.&& &.&.\\
0 & 0 & . &.& .& N/2-1\\
\end{pmatrix},
\end{equation}
whereas the quantum mechanical momentum operators $\hat{p}_{x}$ and $\hat{p}_{y}$ are denoted as $P_{X}$ and $P_{Y}$, respectively, are also represented by $N\times N$ Hermitian matrices which are obtained from the centered Fourier transform of the position matrices $X$ and $Y$ as follows,
\begin{equation}
P_X=F^{-1} X F,\,\,\,P_Y=F^{-1} Y F,
\label{momdef}
\end{equation}
and the centered quantum Fourier transform is given by the $N\times N$ unitary matrix $F$ with matrix elements, 
\begin{equation}
F_{jk}=\frac{1}{\sqrt{N}}\mathrm{exp}\left[\frac{2 \pi i}{N}j k\right],\,\,\, j,k \in [-N/2,N/2-1].
\label{centeredqft}
\end{equation}
Moreover, the $X$ and $P_{X}$ matrices act on the more significant set of $p$ qubits of the computational basis, which is dedicated to the x-axis, and leave the less significant set of $p$ qubits for the y-axis intact. On the other hand, the action of the $Y$ and $P_{Y}$ matrices on the computational basis is just reversed. As an example, 6 qubits are needed for $N=8$, and the most general computational state can be denoted as $\ket{q_5 q_4 q_3 q_2 q_1 q_0}$. The first 3 qubits $q_2$, $q_1$, and $q_0$ are dedicated to the y-axis, and the rest are dedicated to the x-axis. The matrices $X$ and $P_X$ act on the states $\ket{q_5q_4q_3}$ whereas $Y$ and $P_Y$ act on the states $\ket{q_2q_1q_0}$. In addition, $X\, P_{Y}$ and $Y P_{X}$ are basically $X\otimes P_{Y}$ and $P_{X}\otimes Y$, respectively, and can be simplified as follows,
\begin{align}
X\otimes P_Y &=X \otimes (F^{-1}YF)= (I\otimes F^{-1}) (X\otimes Y) (I \otimes F)=(I\otimes F^{-1}).D.(I \otimes F),\nonumber\\
P_X\otimes Y &= (F^{-1}XF) \otimes Y= (F^{-1}\otimes I) (X\otimes Y) (F \otimes I)=(F^{-1}\otimes I).D.(F \otimes I),\nonumber
\end{align}
where we denote, $D=X\otimes Y$.

\subsection{Time evolution operator representation}\label{timeevolution}
In terms of the finite-dimensional position operators $X$, $Y$ and momentum operators $P_{X}$ and $P_{Y}$, the quantum mechanical Hamiltonian in Eq. (\ref{commhamiltonian}) associated with the noncommutative Hamiltonian of Eq. (\ref{eq:NC-Hamiltonian}) is expressed as
\begin{equation}
H=\frac{1}{2M_1}P_X^2+\frac{1}{2M_2}P_Y^2+\frac{1}{2}M_1\Omega_1^2X^2+\frac{1}{2}M_2\Omega_2^2Y^2-l_1X\,P_Y+l_2Y\,P_X\, =\, H_X+H_Y+H_{XY},\label{mathamil}
\end{equation}
where for notational convenience, we denote
\begin{equation}
H_X=\frac{1}{2M_1}P_X^2+\frac{1}{2}M_1\Omega_1^2X^2,\,\,\,
H_Y=\frac{1}{2M_2}P_Y^2+\frac{1}{2}M_2\Omega_2^2Y^2,\,\,\,
H_{XY}=-l_1XP_Y+l_2YP_X.
\nonumber
\end{equation}

The time evolution operator $U(t)$ corresponding to the Hamiltonian in Eq. (\ref{mathamil}) is given by
\begin{equation}
U(t)=e^{-iHt/\hbar}
=e^{-i(H_X+H_Y+H_{XY})t/\hbar}.
\label{unitary}
\end{equation}
Now focusing only on the isotropic case $\omega_{1}=\omega_{2}$, for which $M_{1}=M_{2}=M$, $\Omega_{1}=\Omega_{2}=\Omega$ and $l_{1}=l_{2}=l$, we use the Trotter-Suzuki formula \cite{trotter,suzuki1,suzuki2,suzuki3,suzuki4,suzuki5} to approximate the time evolution operator,
\begin{equation}
e^{-iHt/\hbar}\approx e^{-i\frac{P_X^2}{2M}t/\hbar}e^{-i\frac{1}{2}M\Omega^2X^2t/\hbar} e^{-i\frac{P_Y^2}{2M}t/\hbar}e^{-i\frac{1}{2}M\Omega^2Y^2t/\hbar}
e^{i\,l X P_Y t/\hbar}e^{-i\,l Y P_X t/\hbar}.\label{approxhamil}
\end{equation}
We take time steps of $\delta t=t/n$ so that
\begin{equation}
U(t)=e^{-iHt/\hbar}=(e^{-iH \delta t/\hbar})^n\\
\approx (e^{-i\frac{P_X^2}{2M}\delta t/\hbar}e^{-i\frac{1}{2}M\Omega^2X^2\delta t/\hbar} e^{-i\frac{P_Y^2}{2M}\delta t/\hbar}e^{-i\frac{1}{2}M\Omega^2Y^2\delta t/\hbar}
e^{i\,l X P_Y \delta t/\hbar}e^{-i\,l Y P_X \delta t/\hbar})^n.
\label{trotter}
\end{equation}
For convenience we denote, $U_{X}= e^{-i\frac{1}{2}M\Omega^2X^2\delta t/\hbar}$ and $U_{Y}= e^{-i\frac{1}{2}M\Omega^2Y^2\delta t/\hbar}$. Besides, the unitary operators containing $P_{X}$ and $P_{Y}$, denoted as $U_{P_X}$, $U_{P_Y}$, $U_{XP_Y}$ and $U_{YP_X}$, are further simplified in the following,
\begin{align}
U_{P_X}=e^{-i\frac{P_X^2}{2M}\delta t/\hbar}&=F^{-1} e^{-i\frac{X^2}{2M}\delta t/\hbar} F=F^{-1}\tilde{U}_{X}F,\,\,\,
U_{P_Y}=e^{-i\frac{P_Y^2}{2M}\delta t/\hbar}=F^{-1} e^{-i\frac{Y^2}{2M}\delta t/\hbar} F=F^{-1}\tilde{U}_{Y}F,\label{Upx_Upy}\\
U_{XP_Y}&= e^{i\, l X P_Y \delta t / \hbar}=(I\otimes F^{-1}).e^{i\,l D \delta t / \hbar}. (I\otimes F)=(I\otimes F^{-1}).U^{+}_{XY}. (I\otimes F),\label{Uxpy}\\
U_{YP_X}&= e^{-i\, l Y P_X \delta t / \hbar}=(F^{-1}\otimes I).e^{-i\, l D \delta t / \hbar}.(F\otimes I)=(F^{-1}\otimes I).U^{-}_{XY}. (F\otimes I),\label{Uypx}
\end{align}
where we also denote, $\tilde{U}_{X}= e^{-i\frac{X^2}{2M}\delta t/\hbar}$, $\tilde{U}_{Y}= e^{-i\frac{Y^2}{2M}\delta t/\hbar}$ and $U^{\pm}_{XY}= e^{\pm i\,l D \delta t / \hbar}$. 
Therefore, the time evolution operator is decomposed into discrete Fourier transform operators and diagonal matrices.

\subsection{Execution and interpretation}\label{execution}
To carry out the quantum simulation, we write and execute a quantum program via the Qasm Simulator of Qiskit (version 0.34.0) \cite{Qiskit}. Initially, all the qubits are in $\ket{0}$ state, and we have to prepare the initial state of our simulation. As a starting point, our initial states are the position eigenstate, $|x,y\rangle$, which can be easily prepared by applying NOT gates to the appropriate qubits according to the mapping given in section \ref{compbasis}. After preparing the initial state, we need to implement time evolution according to Eq. (\ref{trotter}). Before executing, the unitary operators of the single trotter step in the time evolution are transpiled into quantum gates via the built-in transpilers of Qiskit, and we use a loop to implement single trotter steps $n$ times for describing the evolution over the chosen time interval. We obtain the final state after implementing the time evolution. The qubits in the final state are measured to obtain counts corresponding to each computational state. Again, by using the inverse map from computational basis to position eigenbasis given in section \ref{compbasis}, we assign the number of counts to the corresponding position eigenstate. In this way, we can find a probability distribution on the mesh after the time evolution of the position eigenstate $|x,y\rangle$. Also, one can easily determine the overlapping of the final state with another position eigenstate by noting the number of counts to the corresponding position eigenstate. Alternatively, the overlapping between the final state and any position eigenstate can be calculated using the CircuitStateFn class of Qiskit Terra, which is essentially an inner product calculation between the two state vectors. In the following sections, we discuss the quantum circuits that can be used to implement the quantum algorithm. Moreover, all circuit diagrams in the subsequent sections are drawn with Qiskit.

\subsubsection{Filter Circuit to Implement Diagonal Unitary Matrices}\label{filtercircuit}
The diagonal unitary matrices can be implemented using filter circuits. We can implement the filter circuit using a multi-controlled NOT gate and only one ancilla qubit instead of multiple Toffoli gates as done in \cite{jain2021quantum}. Filter circuits select a certain state and apply a phase to that state. Using a filter circuit for each state, we can implement an operator corresponding to a diagonal unitary matrix. In our simulation, the exponential of $X^2$, $Y^2$, and $D$ matrices being diagonal unitary matrices are implemented in this manner.
\begin{figure}[H]
    \centering
    \includegraphics[width=5.7cm]{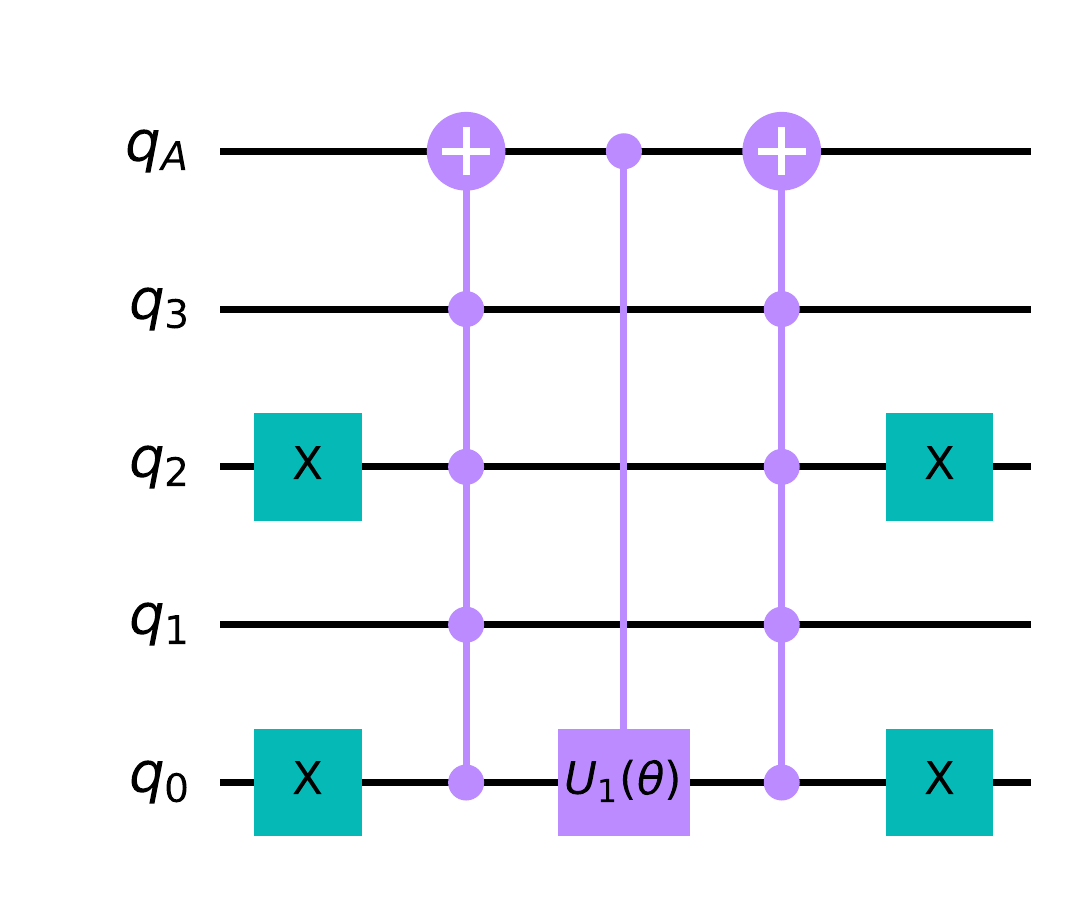}
    \caption{Filter Circuit for the State $\ket{1010}$}
    \label{filter}
\end{figure}
Let us demonstrate the action of a filter circuit on a state $|1010\rangle$ as shown in Fig. \ref{filter}. The multi-controlled NOT gate acting on the ancilla qubit $q_A$ is activated only when all of its control qubits are in $\ket{1}$ state. Therefore, in the case of $|1010\rangle$, first, the NOT gates invert $q_2$ and $q_0$ to $\ket{1}$ and activate the adjacent multi-controlled NOT gate. Now the ancilla qubit $q_A$, initially in state $\ket{0}$, switches to $\ket{1}$ due to the activation of the multi-controlled NOT gate. At this point, since $q_A$ is in state $\ket{1}$, the controlled $U_1$($\theta$) gate will be activated. It will act on $q_0$ and add the phase $e^{i\theta}$ to the circuit. After that, the rightmost multi-controlled NOT gate will switch the ancilla qubit $q_A$ from $\ket{1}$ to $\ket{0}$, and finally, the two NOT gates at the right act on $q_2$ and $q_0$ to invert them from $\ket{1}$ to $\ket{0}$. Hence, the state $\ket{1010}$ is transformed to $e^{i\theta}\ket{1010}$ by the action of the filter circuit. Note that this filter circuit does not alter any state other than $\ket{1010}$. After the operation of the filter circuit, the ancilla qubit $q_A$ is in $\ket{0}$ state. We can append another filter circuit with a new combination of NOT gates to select another state. We can also choose another value of the applied phase $\theta$ by changing the controlled $U_1$ gate parameter. Consequently,  to implement a diagonal unitary matrix, we apply filter circuits for each of the states with the phase $e^{i\theta_{i}}$ corresponding to its $i$-th diagonal entry. 
\subsubsection{Implementation of Centered Quantum Fourier Transform}
\begin{figure}[h!]
    \centering
    \includegraphics[width=14cm]{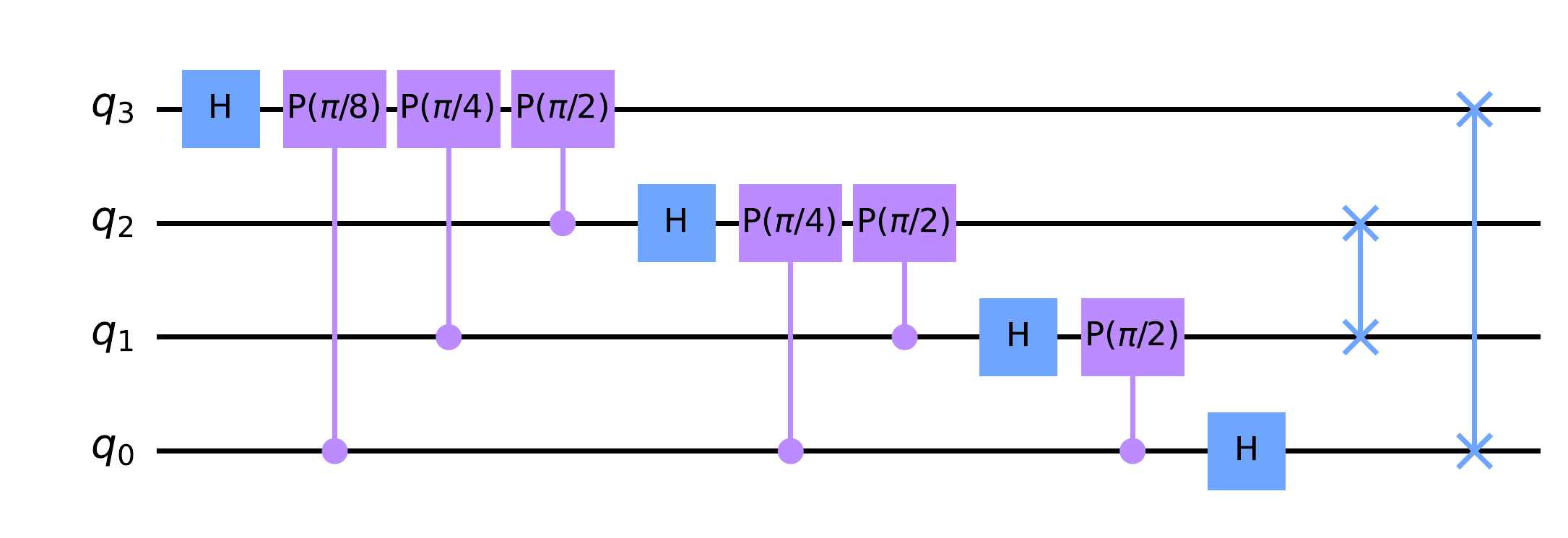}
    \caption{QFT circuit for Four Qubits}
    \label{qftfig}
\end{figure}

\begin{figure}[h!]
    \centering
    \includegraphics[width=5cm]{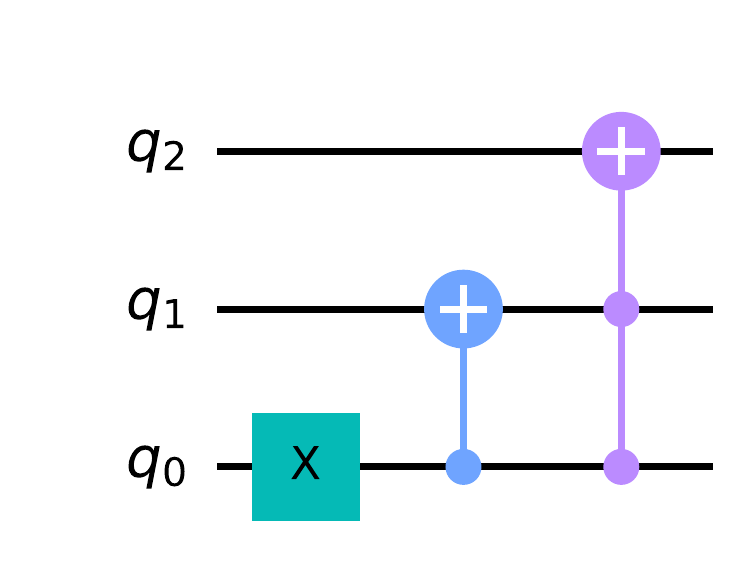}
    \caption{Permutation circuit for 3 qubits}
    \label{perm}
\end{figure}

As the momentum operators are defined using centered discrete Fourier transformation as in Eq. (\ref{momdef}), we implement it using the quantum circuit for the quantum Fourier transform (QFT) discussed in Qiskit textbook \cite{qft} and a permutation circuit, explained below, to make it centered. In Fig. \ref{qftfig}, we only present the circuit diagram for QFT acting on four qubits. Moreover, the inverse QFT can be implemented by applying the gates in reverse order since it is a unitary operation.
In our implementation, the centered discrete Fourier transform, defined as $F=P^{N/2} F^{d} P^{-N/2}$, is obtained by the permutation of the elements of the standard discrete Fourier transform $F^{d}$ which is defined as $F^{d}_{j k}=\frac{1}{\sqrt{N}}\mathrm{exp}\left[\frac{2 \pi i}{N}jk\right]$ where $j,k=0,1,...,N-1$ \cite{somma2015quantum}. Therefore, we need to define a quantum circuit that permutes the computational basis states. The matrix representations of the permutation denoted as $P$ and its inverse $P^{-1}$ are given by
\begin{equation}
P =
\begin{pmatrix}
0 & 1 & 0& . &  .& .&0\\
0 & 0 & 1& . & . & .&0 \\
. & . & & 1 &  & &. \\
.&.&& &.& &.\\
.&.&& &&. &.\\
.&.&& && &1\\
1 & 0 & 0 &.& .& . &0\\
\end{pmatrix},\hspace{5\baselineskip}
P^{-1} =
\begin{pmatrix}
0 & 0 & 0& . &  .& .&1\\
1 & 0 & 0& . & . & .&0 \\
. & 1 & & . &  & &. \\
.&.&1& && &.\\
.&.&&. && &.\\
.&.&& &.& &\\
0 & 0 & 0 &.& .& 1 &0\\
\end{pmatrix}.\hspace{\baselineskip}
\label{permat}
\end{equation}
Let us note how the operator $P$ maps the computational basis states appropriately for a case of three qubits from Table \ref{permtable}.
\begin{table}[h!]
    \centering
\begin{tabular}{|p{1cm}|p{1cm}|p{1cm}|p{1cm}|p{1cm}|p{1cm}|  }
 \hline
 \multicolumn{3}{|c|}{$\ket{\psi}$}& \multicolumn{3}{c|}{$P\ket{\psi}$}\\
  \hline
  $q_2$ & $q_1$ & $q_0$ &$q_2$ & $q_1$ & $q_0$ \\
 \hline
  \cellcolor{lightgray}$\ket{0}$& \cellcolor{lightgray}$\ket{0}$  &$\ket{0}$& \cellcolor{lightgray}$\ket{1}$& \cellcolor{lightgray}$\ket{1}$&$\ket{1}$\\
  $\ket{0}$&$\ket{0}$&$\ket{1}$& $\ket{0}$&$\ket{0}$&$\ket{0}$\\
  $\ket{0}$&\cellcolor{lightgray}$\ket{1}$&$\ket{0}$&$\ket{0}$ &\cellcolor{lightgray}$\ket{0}$&$\ket{1}$\\
  $\ket{0}$&$\ket{1}$&$\ket{1}$& $\ket{0}$&$\ket{1}$&$\ket{0}$\\
  \cellcolor{lightgray}$\ket{1}$&\cellcolor{lightgray}$\ket{0}$&$\ket{0}$& \cellcolor{lightgray}$\ket{0}$&\cellcolor{lightgray}$\ket{1}$&$\ket{1}$\\
  $\ket{1}$&$\ket{0}$&$\ket{1}$& $\ket{1}$&$\ket{0}$&$\ket{0}$\\
  $\ket{1}$&\cellcolor{lightgray}$\ket{1}$&$\ket{0}$& $\ket{1}$&\cellcolor{lightgray}$\ket{0}$&$\ket{1}$\\
  $\ket{1}$&$\ket{1}$&$\ket{1}$& $\ket{1}$&$\ket{1}$&$\ket{0}$\\
 \hline
\end{tabular}
\caption{Permutation Operation on 3 qubit states}
\label{permtable}
\end{table} First of all, the $q_0$ is inverted under the operation of $P$, i.e., if it is initially in the state $\ket{0}$, it gets mapped to $\ket{1}$ and vice versa. Now, $q_1$ is inverted only when $q_0$ is in $\ket{0}$ initially. Moreover, $q_2$ is inverted only when both $q_1$ and $q_0$ are initially in $\ket{0}$. Both of these cases are highlighted in Table \ref{permtable}. Furthermore, this operation can be extended for an arbitrary number of qubit states. In Fig. \ref{perm}, we present a representative quantum circuit that can carry out the permutation operation by applying a NOT gate to the first qubit and successive controlled-NOT gates on the respective qubits. Here, the NOT gate acting on the first qubit $q_0$ inverts the first qubit. After that, the $q_1$ will be inverted due to the activation of the CNOT gate. Finally, when $q_0$ and $q_1$ are initially in $\ket{0}$ state, the $q_2$ is inverted by the CCNOT gate. Accordingly, when one has more qubits, the addition of consecutive multi-controlled NOT gates will carry out the permutation operation $P$ on the corresponding states. We can apply the gates in reverse to implement $P^{-1}$ since they implement unitary operations. The centered discrete Fourier transform can be implemented by sandwiching the QFT circuit between $N/2$ repetitions of the permutation circuit and its inverse.
\subsubsection{Implementation of a single Trotter step}
\begin{figure}[H]
    \centering
    \begin{picture}(190,190)(15,-20)% width and height of the picture
    \put(-150,0){\includegraphics[width=\textwidth]{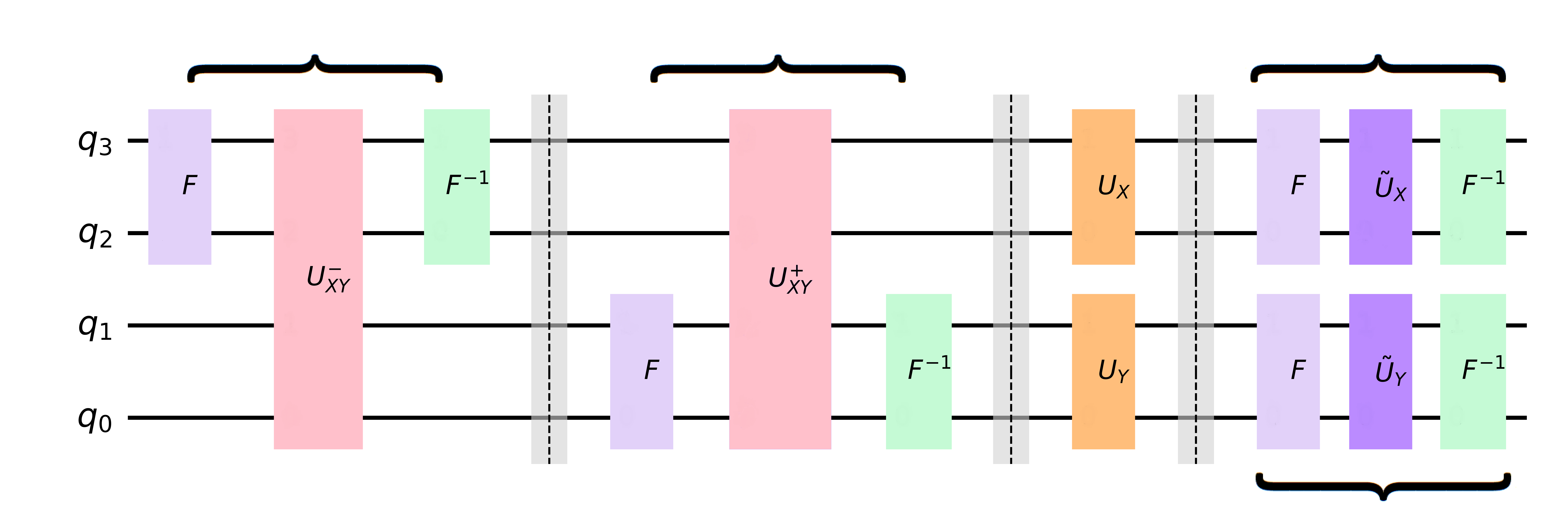}}
    \put(-57,160){\Large$U_{YP_X}$}
    \put(85,160){\Large$U_{XP_Y}$}
    \put(290,160){\Large$U_{P_X}$}
    \put(290,-10){\Large$U_{P_Y}$}
    \end{picture}
    \caption{\centering A schematic diagram of a single Trotter step for 4 qubits.}
    \label{step}
\end{figure}
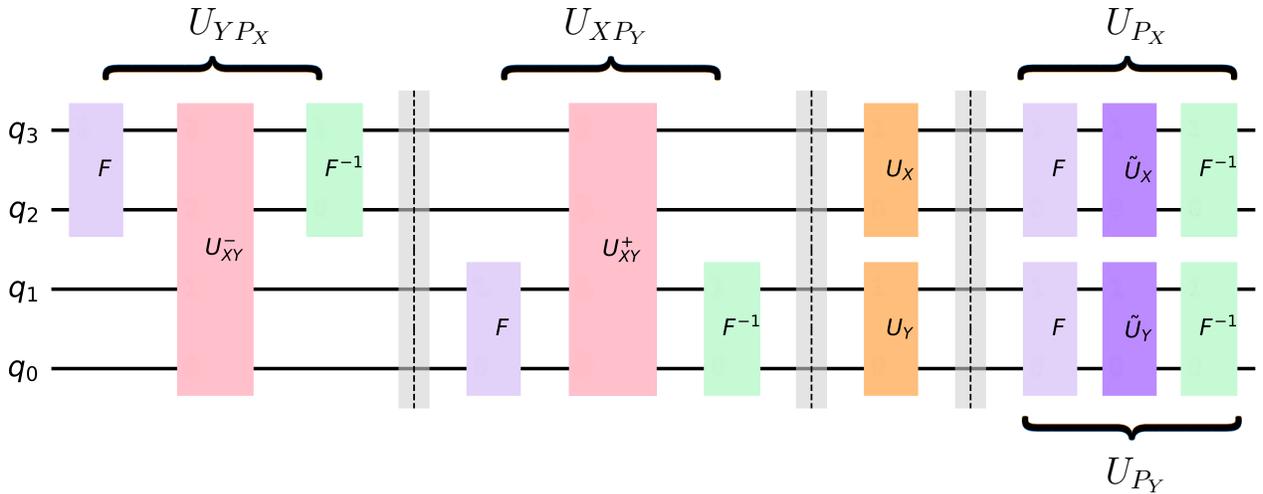
In Fig. \ref{step}, we illustrate the implementation of a single trotter step given in Eq. (\ref{trotter}) on  four qubits which consists of unitary diagonal matrices as defined in section \ref{timeevolution} and $F$ being the centered discrete Fourier transform matrix. Also, an ancilla qubit is required to implement those diagonal unitary matrices via filter circuits which is made implicit in Fig. \ref{step}. Therefore, the time evolution of an initial multi-qubit state, say $|i\rangle$ over time $t$, denoted by $|f\rangle$ can be determined by applying the above-mentioned single Trotter step $n$ successive times, each for $\delta t=t/n$, so that one obtains $|f\rangle = \left[U(\delta t)\right]^{n}|i\rangle$.

\section{Discussion}\label{discussion}
We carry out the quantum simulation of the two-dimensional isotropic harmonic oscillator in zero magnetic field with spatial commutativity following the algorithm stated in section \ref{quantumalgo} and discern the effect of the noncommutativity parameter $\theta$ on the simulation itself. The associated Hamiltonian is,
\begin{equation}
H=\frac{1}{2M}\left(P_X^2+P_Y^2\right)+\frac{1}{2}M\Omega^2\left(X^2+Y^2\right)-l\left(X\,P_Y-Y\,P_X\right),\label{isoham}
\end{equation}
where, the parameters $M$, $\Omega$ and $l$ in terms of parameters of Eq. (\ref{eq:NC-Hamiltonian}) in the isotropic limit, $\omega_{1}=\omega_{2}=\omega$ and setting\footnote{In this work we adopt $\hbar =1$, $c=1$ and the mass is expressed in an arbitrary unit. Also, the length $x$ and time $t$ are both considered in the inverse of the mass unit. Besides, $m\omega\theta$ is dimensionless in this unit system.} $\hbar=1$, $B=0$, $r=0$ and $s=1/2$ in Eq. (\ref{parameters}), are given as
\begin{equation}
    M=\frac{m}{1+\left(\frac{m\theta\omega}{2}\right)^2}\,,\,\,\,\Omega=\omega\sqrt{1+\left(\frac{m\theta\omega}{2}\right)^2},\,\,\, l=\frac{m \theta\omega^{2}}{2}.\label{isoharmpara}
\end{equation}

\begin{figure}[H]\begin{minipage}[c]{0.33\textwidth}\centering\includegraphics[trim={2cm 1cm 1.2cm 4cm},clip,width= \textwidth]{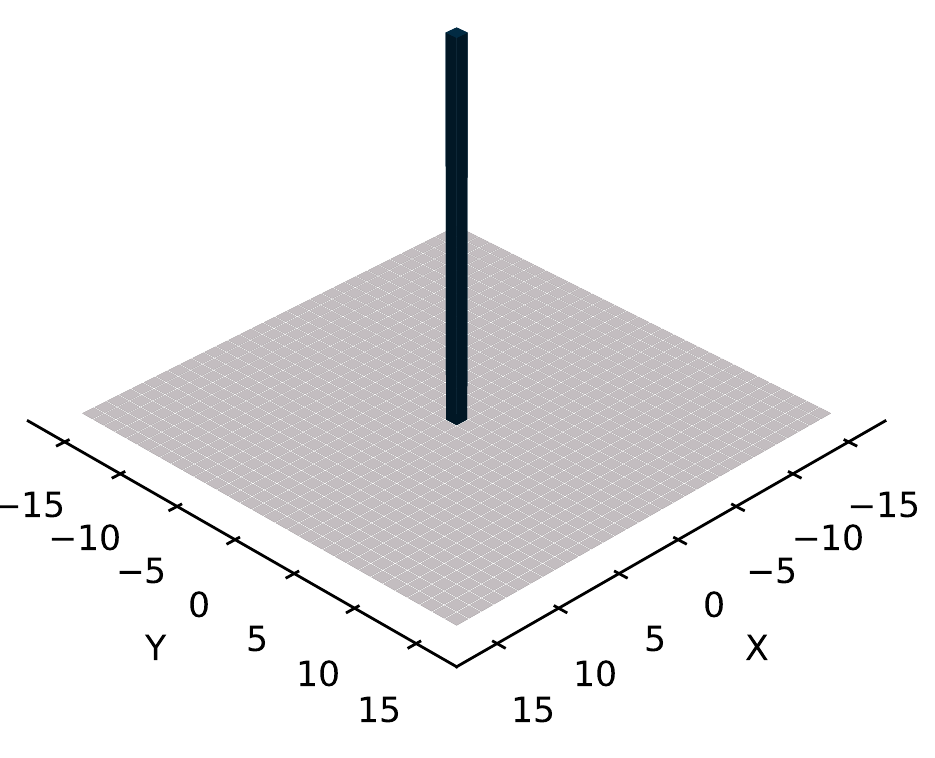}\begin{picture}(0,0)\put(-160,120){\Large{$\theta=0$}}\end{picture}
    \end{minipage}\begin{minipage}[c]{0.33\textwidth}\centering
       \includegraphics[trim={2cm 1cm 1.2cm 4cm},clip,width= \textwidth]{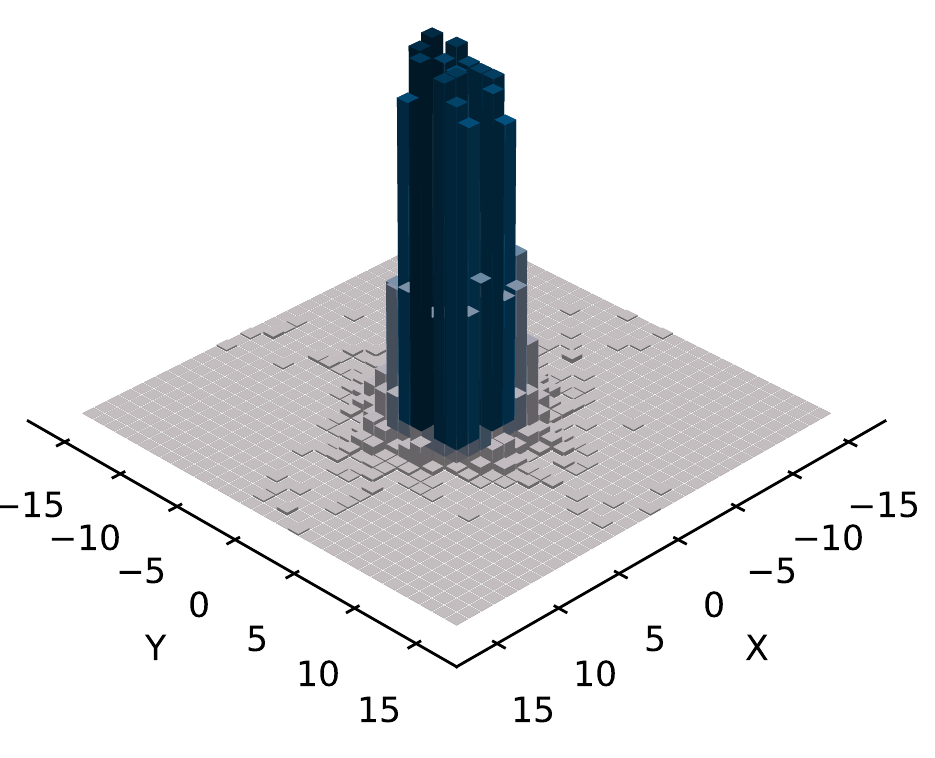}
        \end{minipage}\begin{minipage}[c]{0.33\textwidth}\centering
       \includegraphics[trim={2cm 1cm 1.2cm 4cm},clip,width= \textwidth]{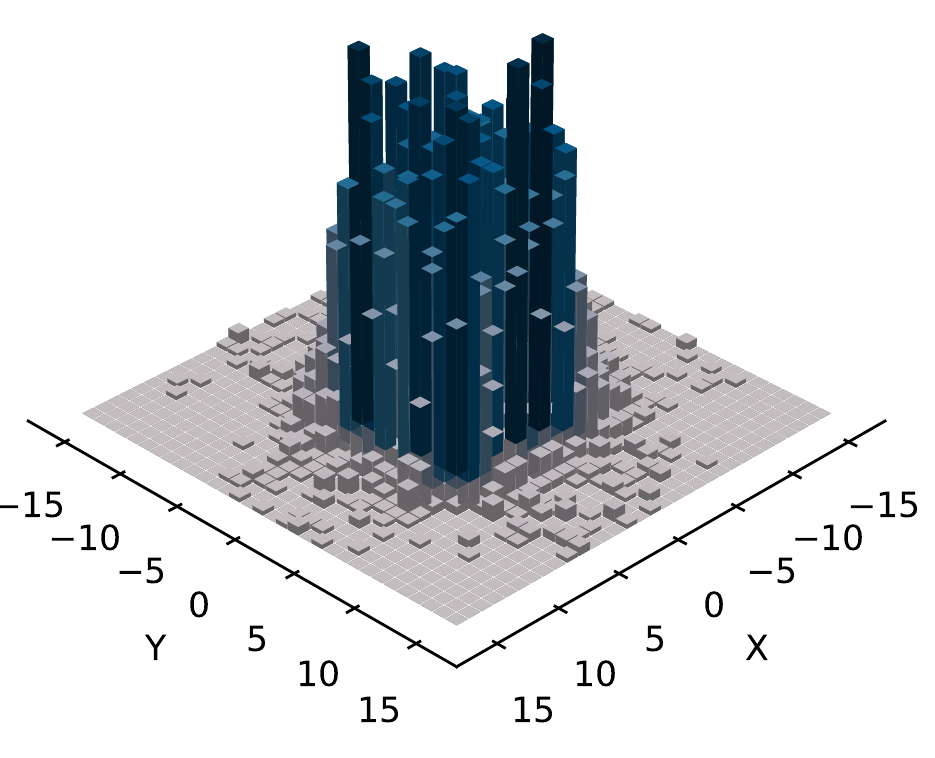}
        \end{minipage}
        \begin{minipage}[c]{0.33\textwidth}\centering
       \includegraphics[trim={2cm 1cm 1.2cm 4cm},clip,width= \textwidth]{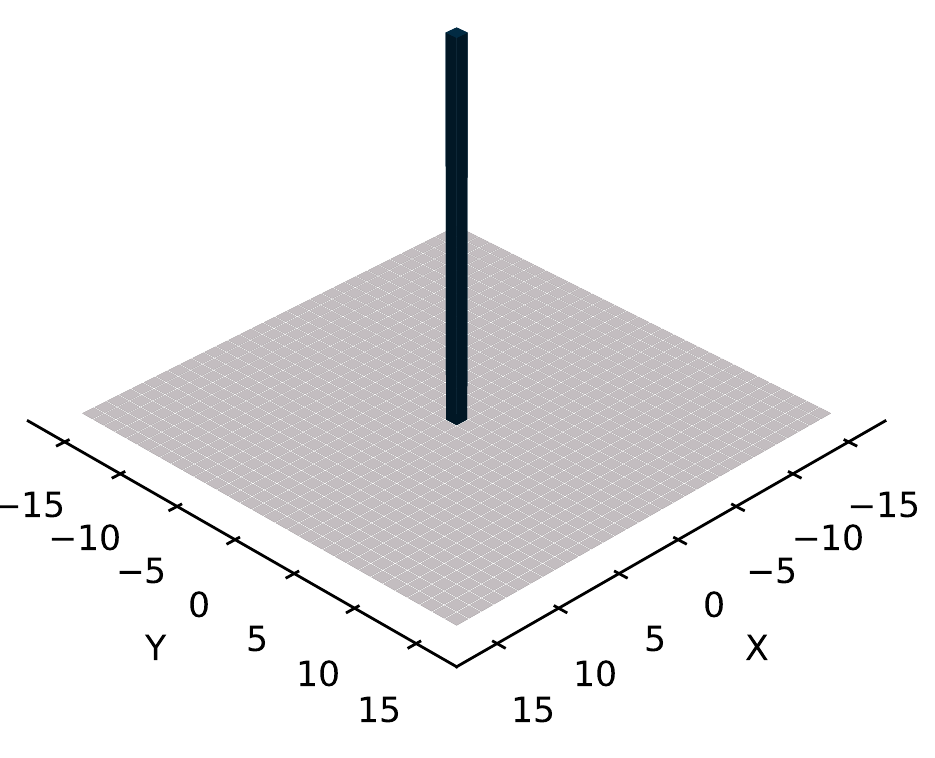}\begin{picture}(0,0)
      \put(-160,120){\Large{$\theta=1$}}
       \end{picture}
       \caption*{$t=0$}
        \end{minipage}\begin{minipage}[c]{0.33\textwidth}\centering
        \includegraphics[trim={2cm 1cm 1.2cm 4cm},clip,width= \textwidth]{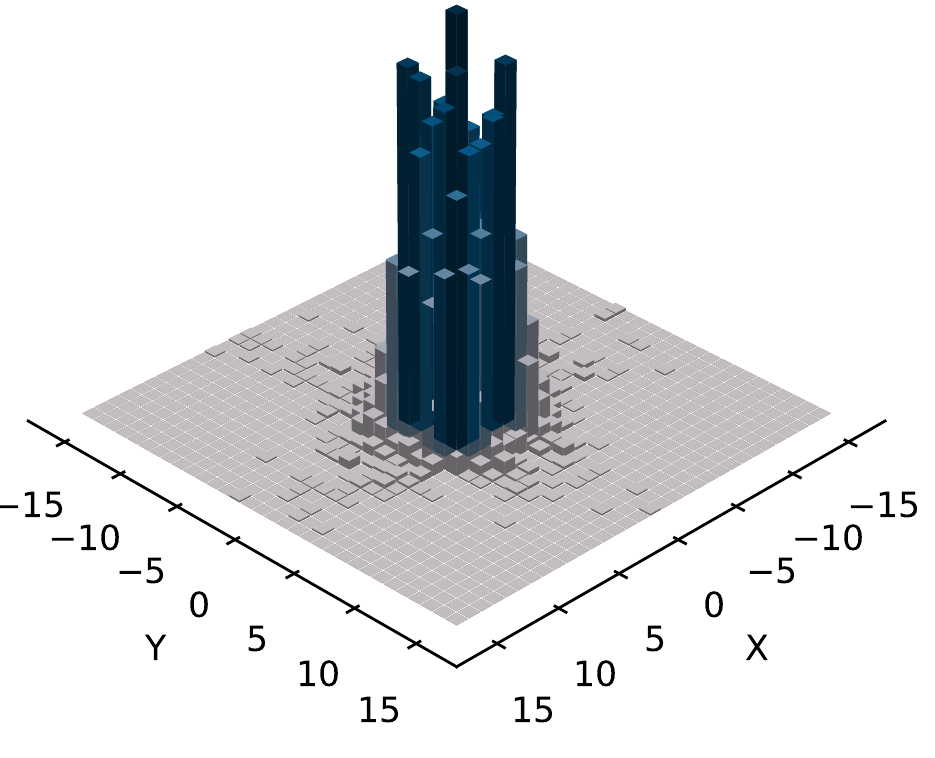}\caption*{$t=0.1$}
        \end{minipage}\begin{minipage}[c]{0.33\textwidth}\centering
        \includegraphics[trim={2cm 1cm 1.2cm 4cm},clip,width= \textwidth]{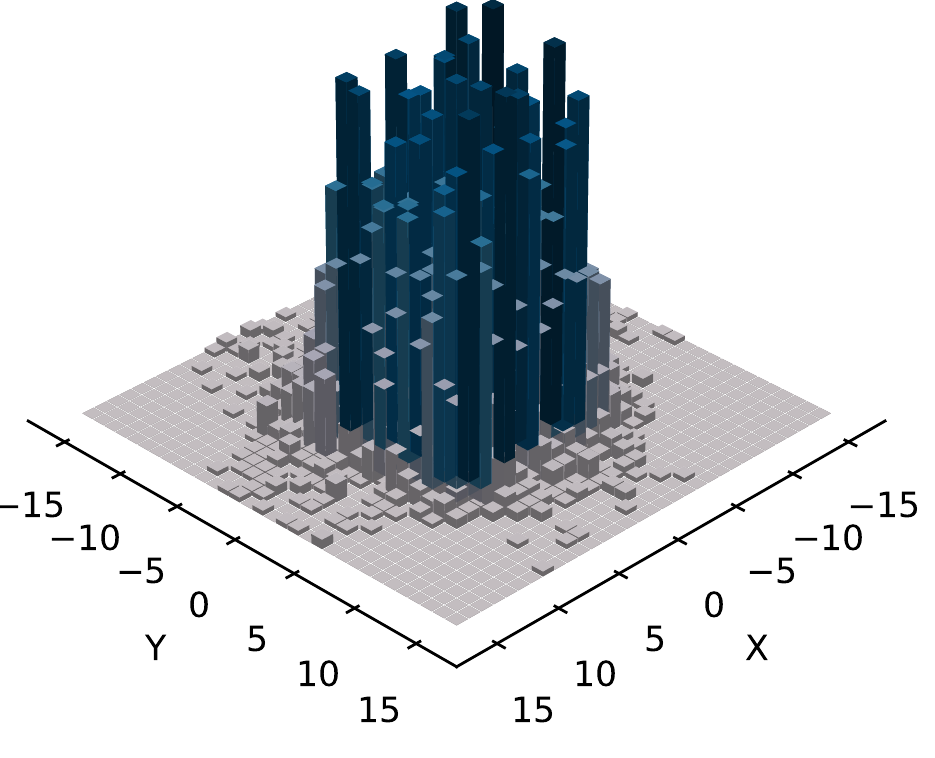}\caption*{$t=0.2$}
        \end{minipage}
        \caption{Probability distribution at $t=0$, 0.1 and 0.2 for $\theta=0$ and $\theta=1$ generated with 8192 shots. The parameter values are $N=32$, $m=0.5$, $\hbar$=1, $\omega$=1, $\delta t=0.02$}
                \label{timefig}
                \end{figure}

\begin{figure}[H] \begin{minipage}[c]{0.5\textwidth}\centering
\includegraphics[width=\linewidth]{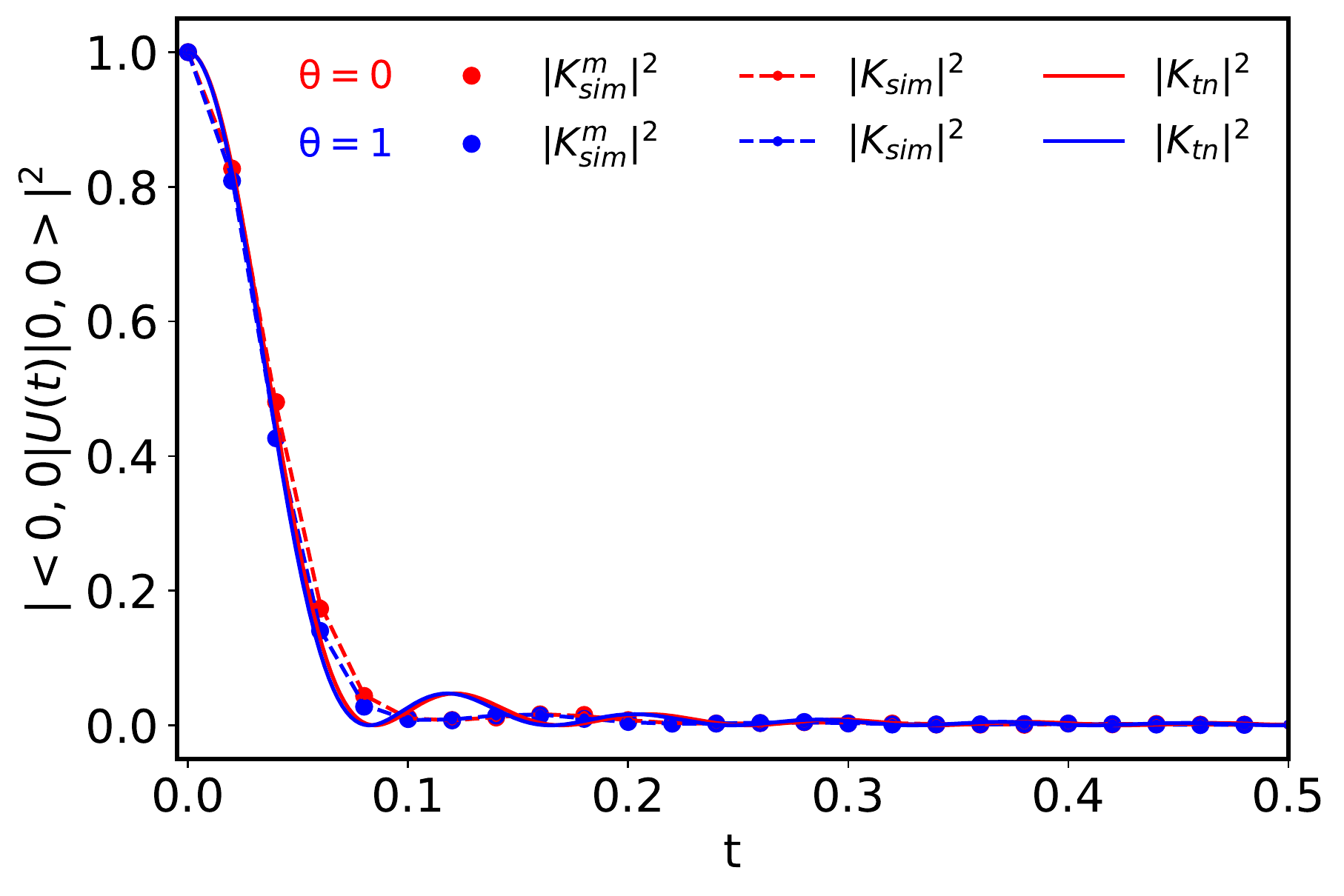}
\begin{picture}(0,0)
\put(-50,43){\includegraphics[width=0.6\linewidth]{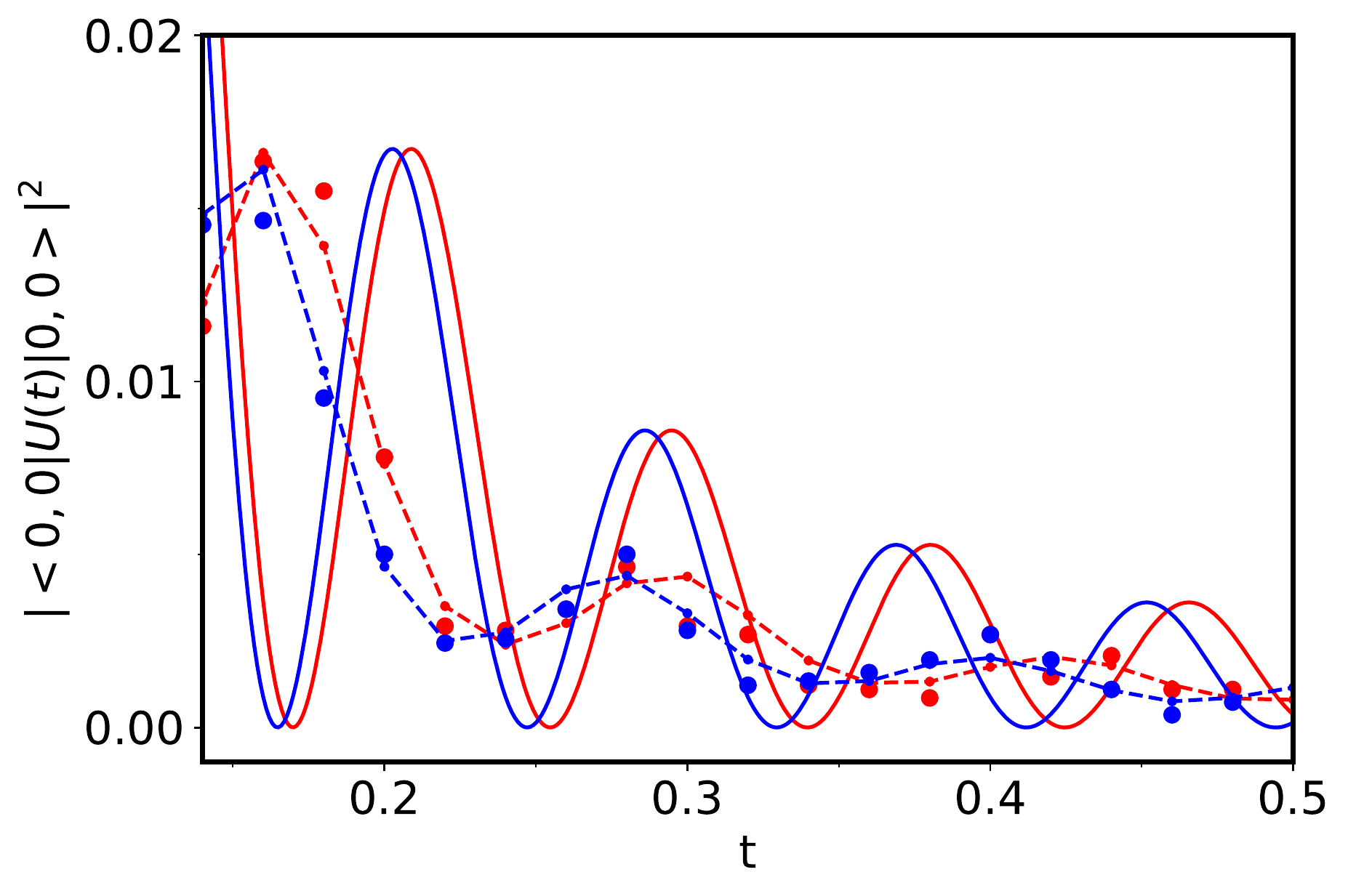}}
\end{picture}
(a)
\end{minipage}
 \begin{minipage}[c]{0.5\textwidth}\centering
\includegraphics[width=\linewidth]{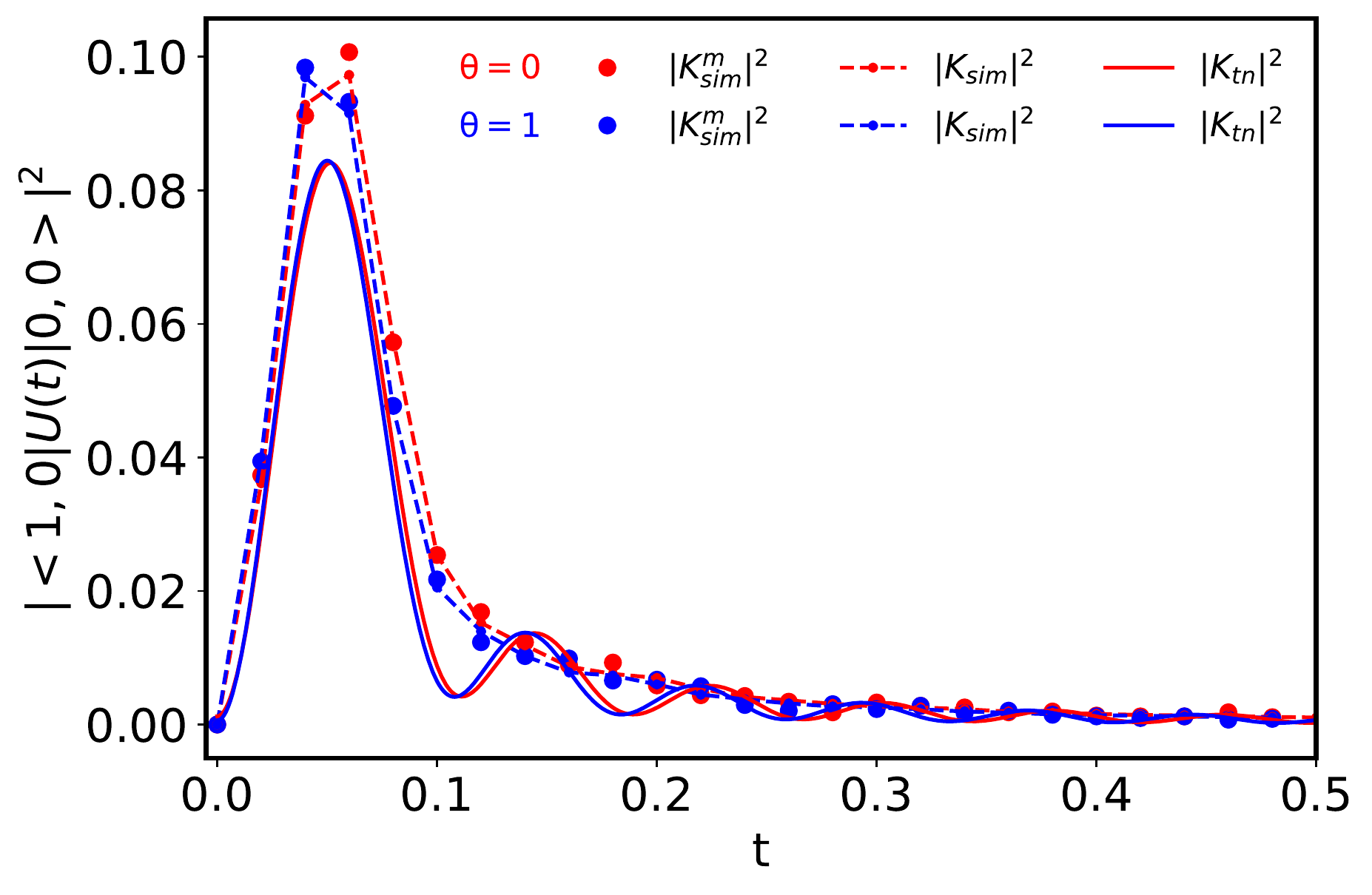}
\begin{picture}(0,0)
\put(-28,50){\includegraphics[width=0.57\linewidth]{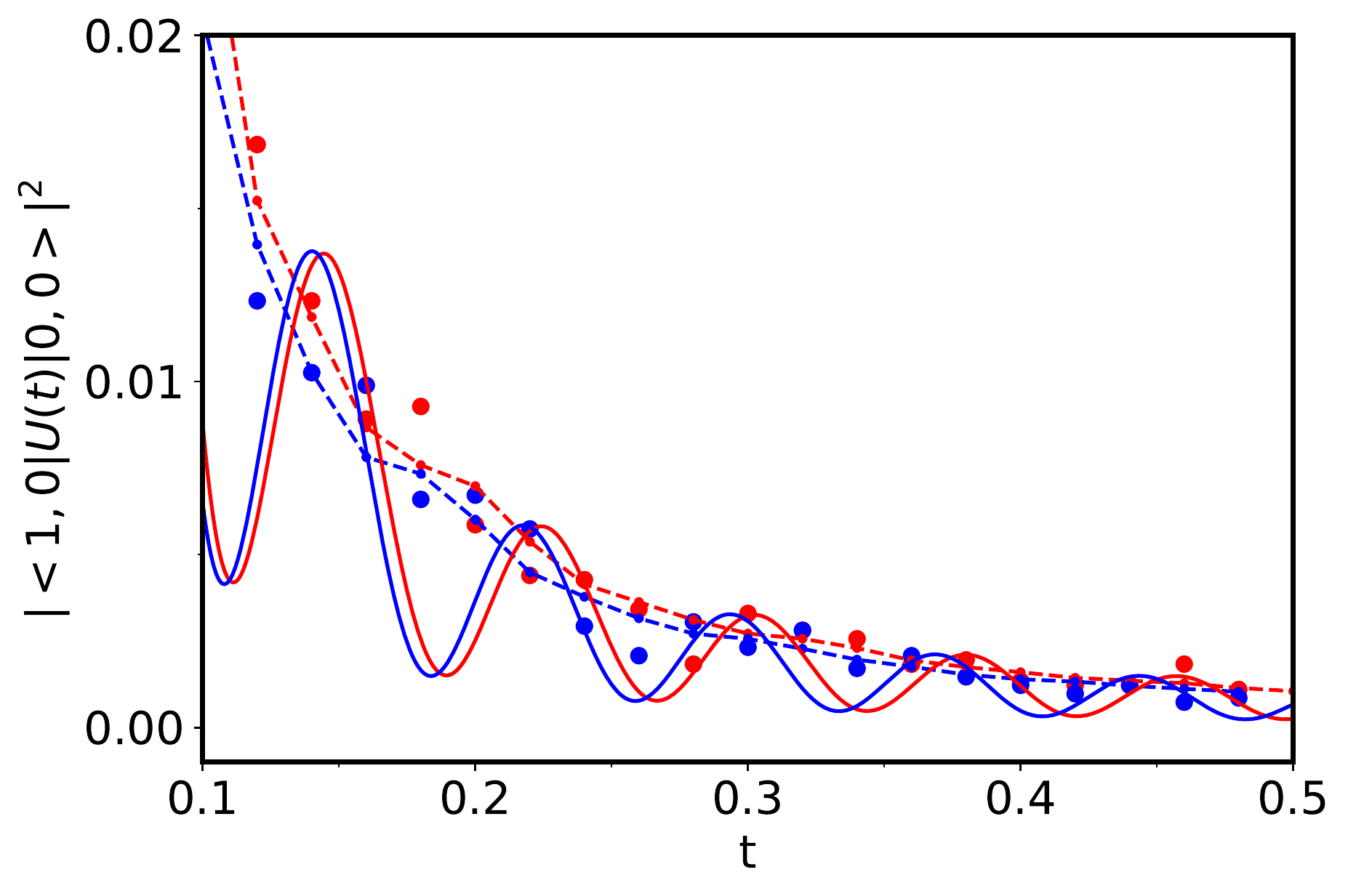}}
\end{picture}
(b)
\end{minipage} 

 \begin{minipage}[c]{0.5\textwidth}\centering \includegraphics[width=\linewidth]{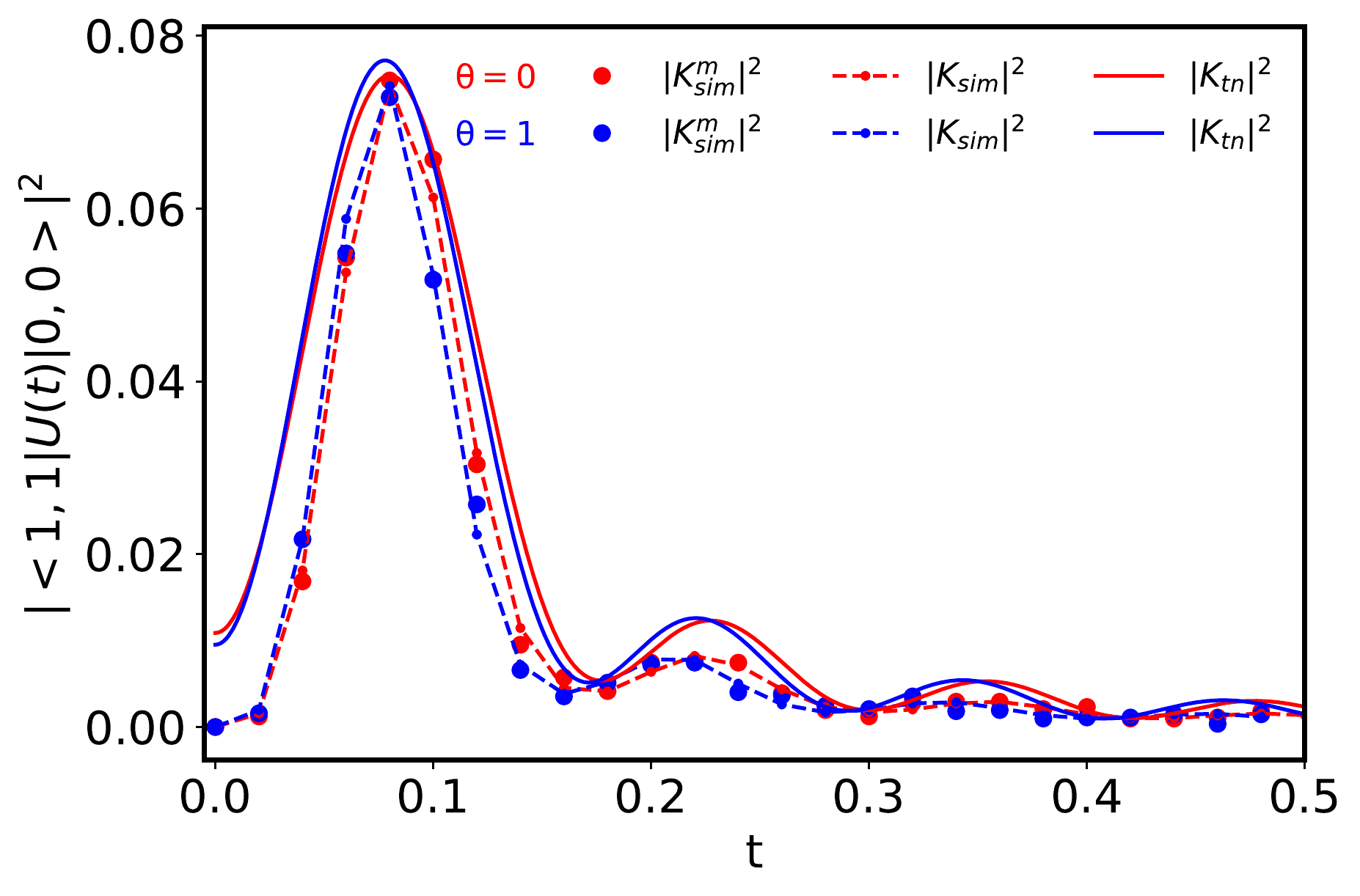}
\begin{picture}(0,0)
\put(-25,55){\includegraphics[width=0.55\linewidth]{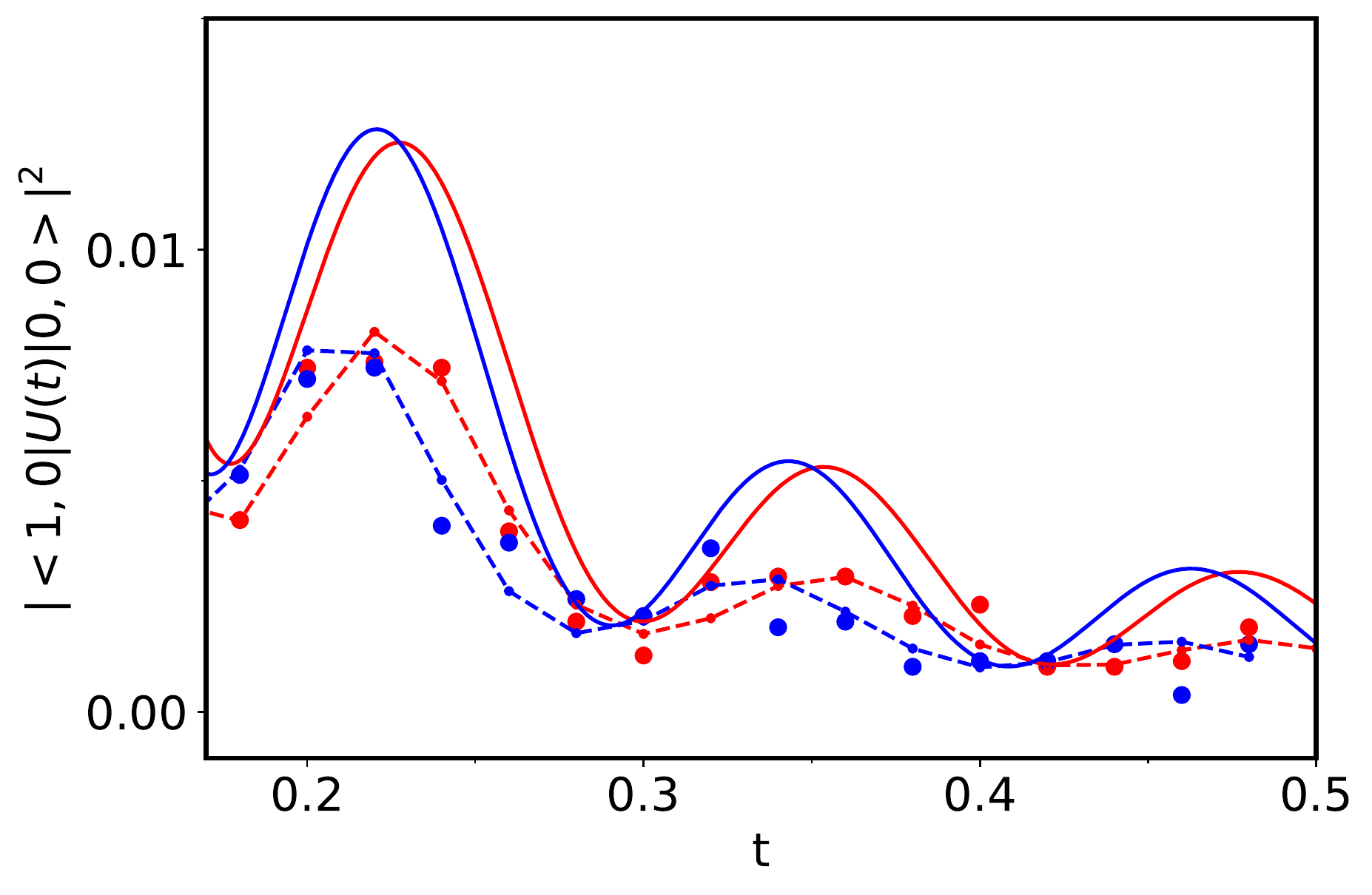}}
\end{picture}
(c)
\end{minipage}
\begin{minipage}[c]{0.5\textwidth}\centering \includegraphics[width=\linewidth]{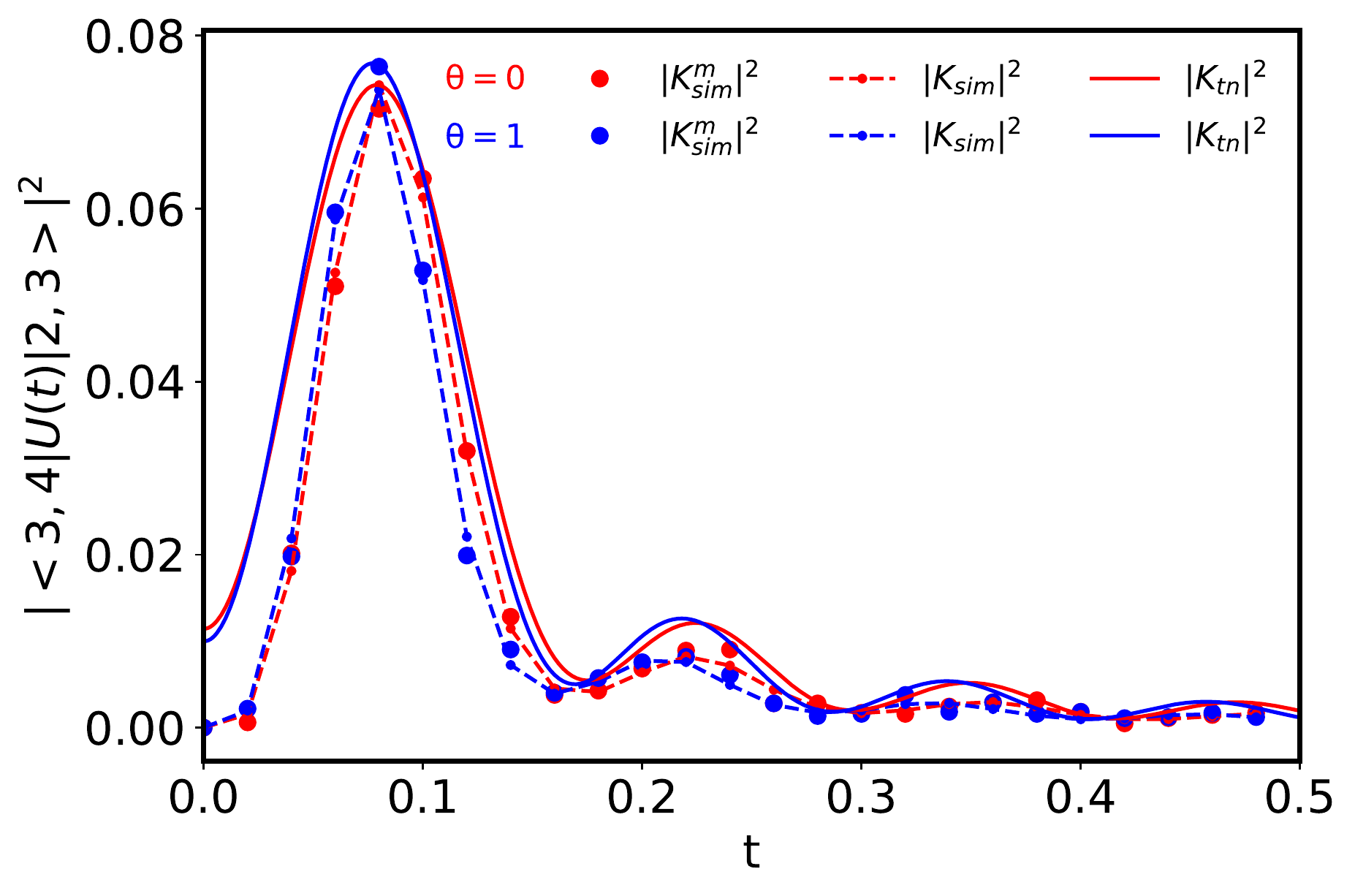}
\begin{picture}(0,0)
\put(-25,55){\includegraphics[width=0.55\linewidth]{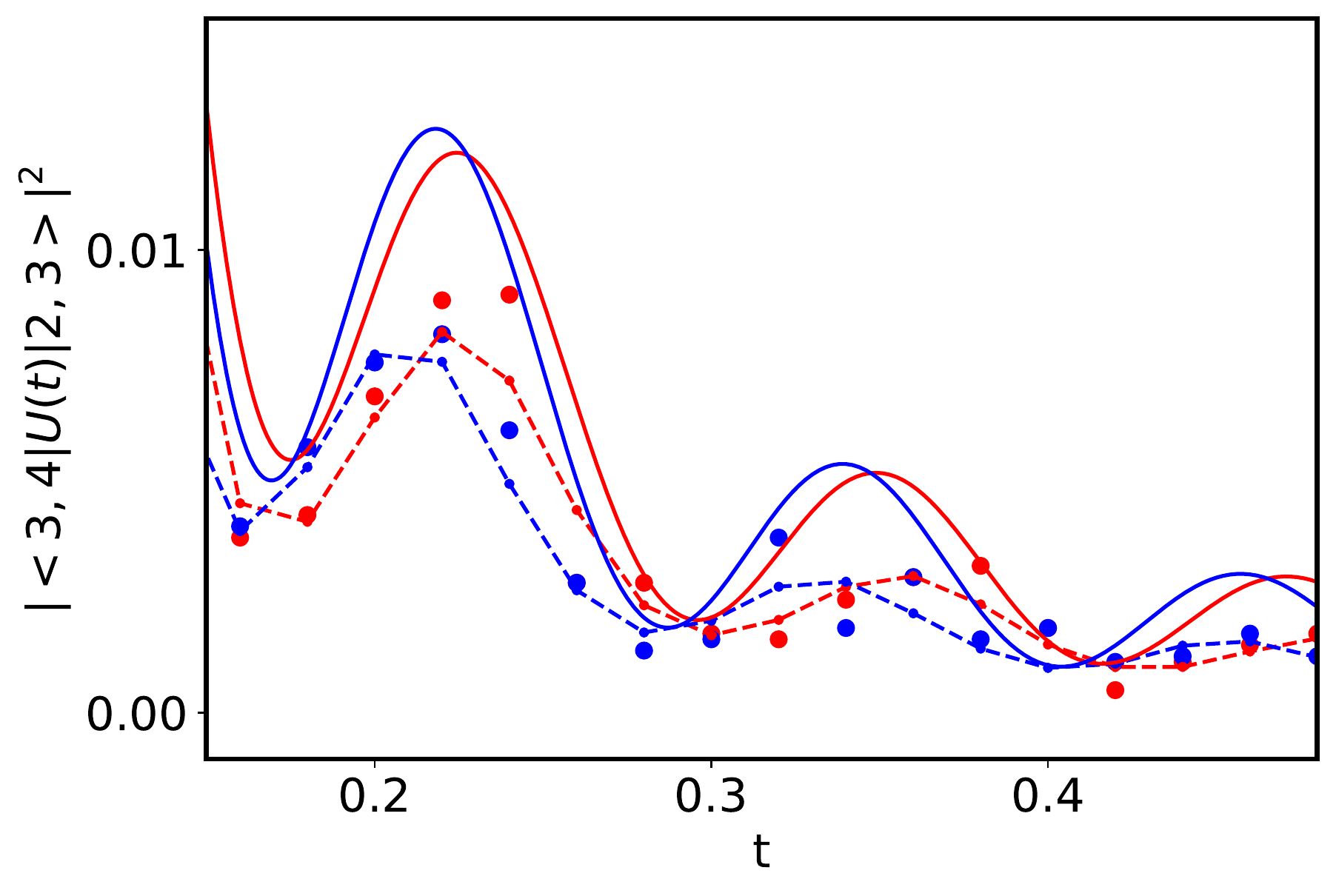}}
\end{picture}
(d)
\end{minipage}
\caption{Variation of the overlapping between two position states, $|\bra{x',y'}U(t)\ket{x,y}|^2$ as a function of time $t$ for noncommutativity parameter, $\theta=0$ and $1$. Here, we consider $N=32$, $m=0.5$, $\omega=1$ and the single Trotter time-step, $\delta t=0.02$. We see that the theoretical values of the overlapping between different position states given by the normalized truncated propagator $|K_{tn}|^{2}$ and those given by the direct computation $|K_{sim}|^{2}$ and the measurement (taking 8192 shots) $|K^{m}_{sim}|^{2}$ after carrying out the simulation over time $t=0.5$ are in good agreement. Additionally, the maximum total energy eigenvalues $n^{\mathrm{max}}_{tot}$ considered in the normalized truncated propagators in figs. (a), (b), (c) and (d) are 72, 78, 50 and 52, respectively. Besides, all of the insets present the zoomed variations of the overlapping for the larger value of time, $t\simgt 0.1$.}  
\label{result1}
\end{figure}

\begin{figure}[H]
 \begin{minipage}[c]{0.5\textwidth}\centering 
  \includegraphics[width=\textwidth]{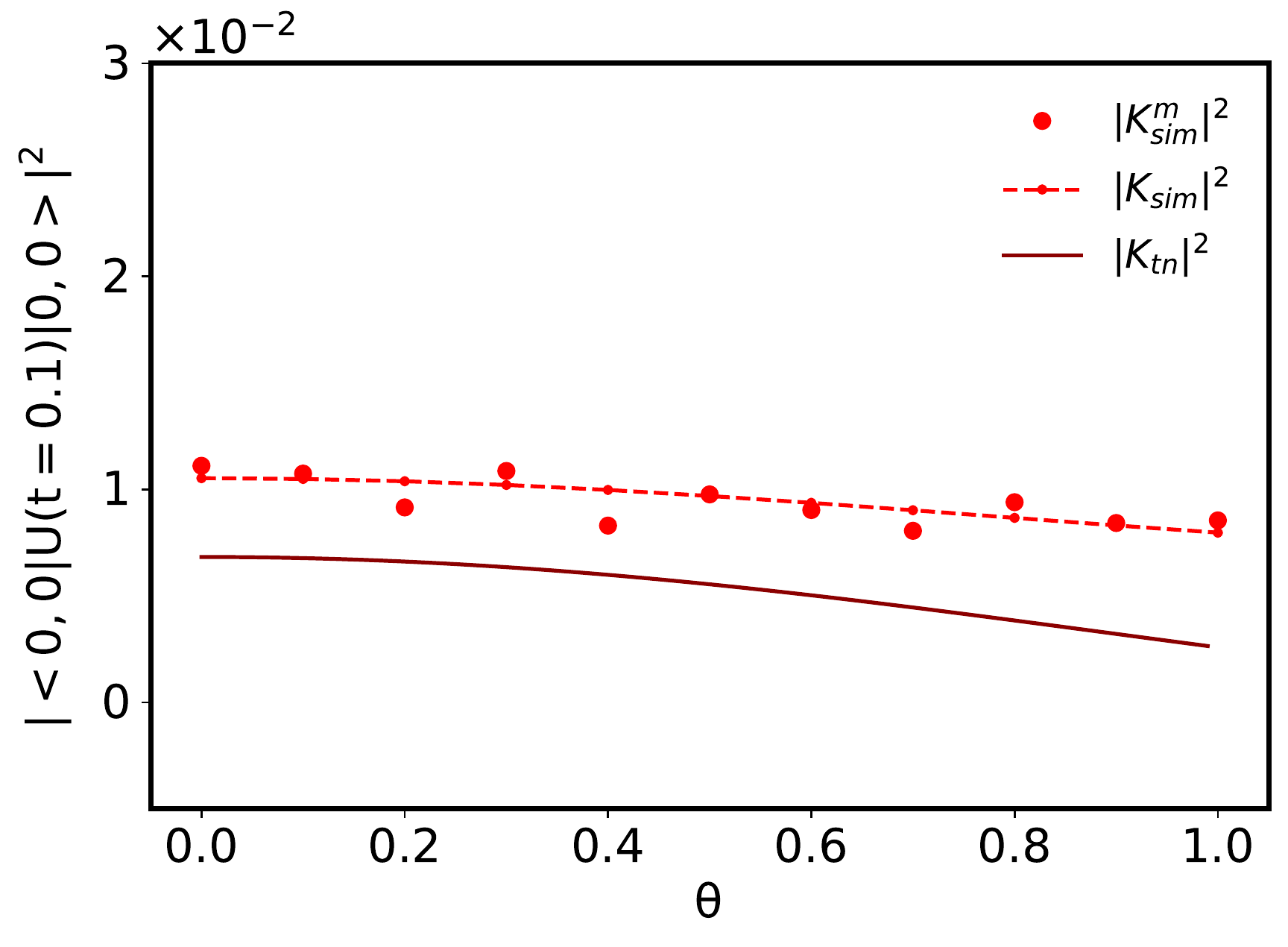}
  (a)
                \end{minipage}
         \begin{minipage}[c]{0.5\textwidth}\centering
         
       \includegraphics[width=\textwidth]{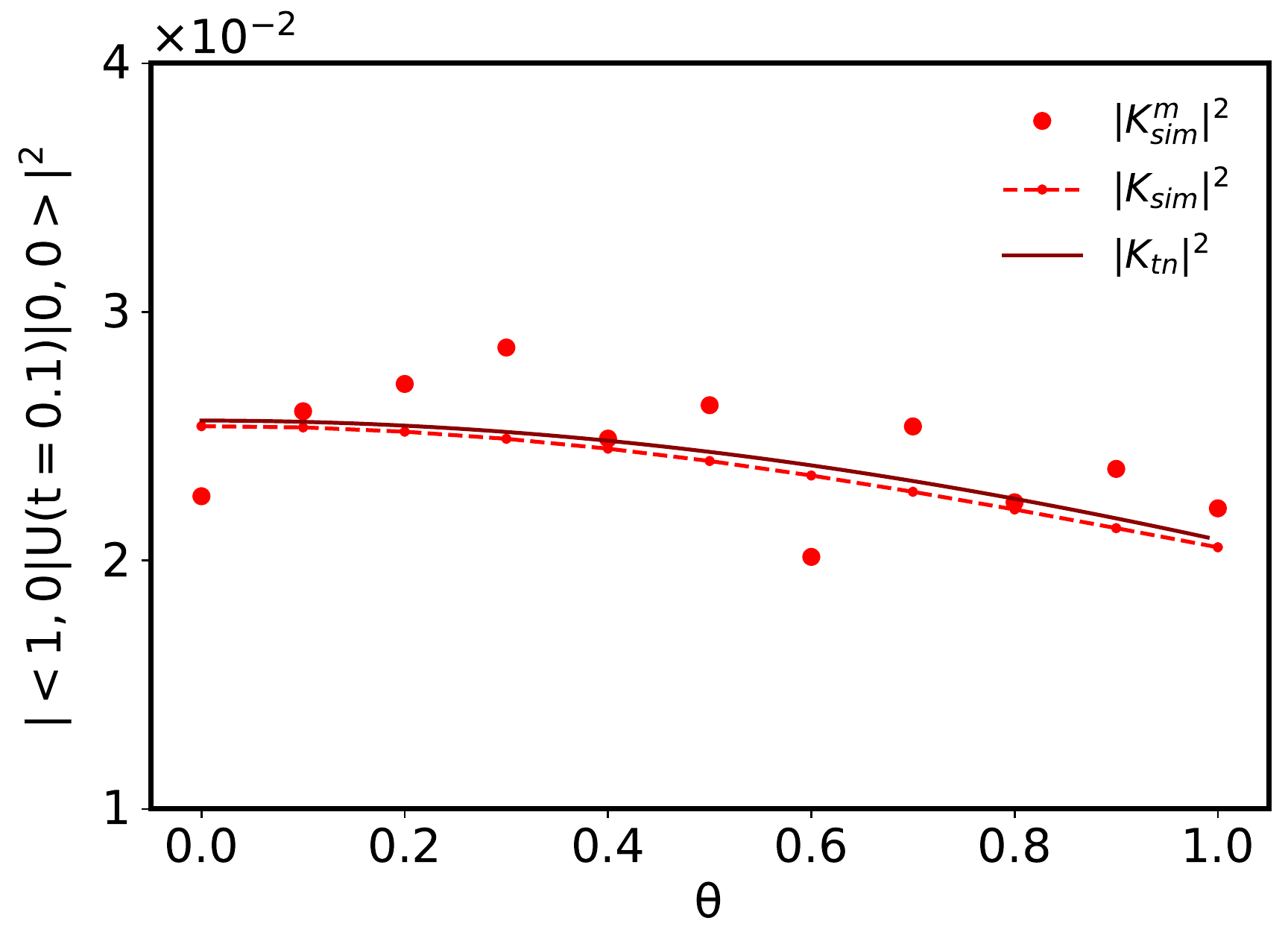}
       (b)
                \end{minipage} \begin{minipage}[c]{0.5\textwidth}\centering 
       \includegraphics[width=\textwidth]{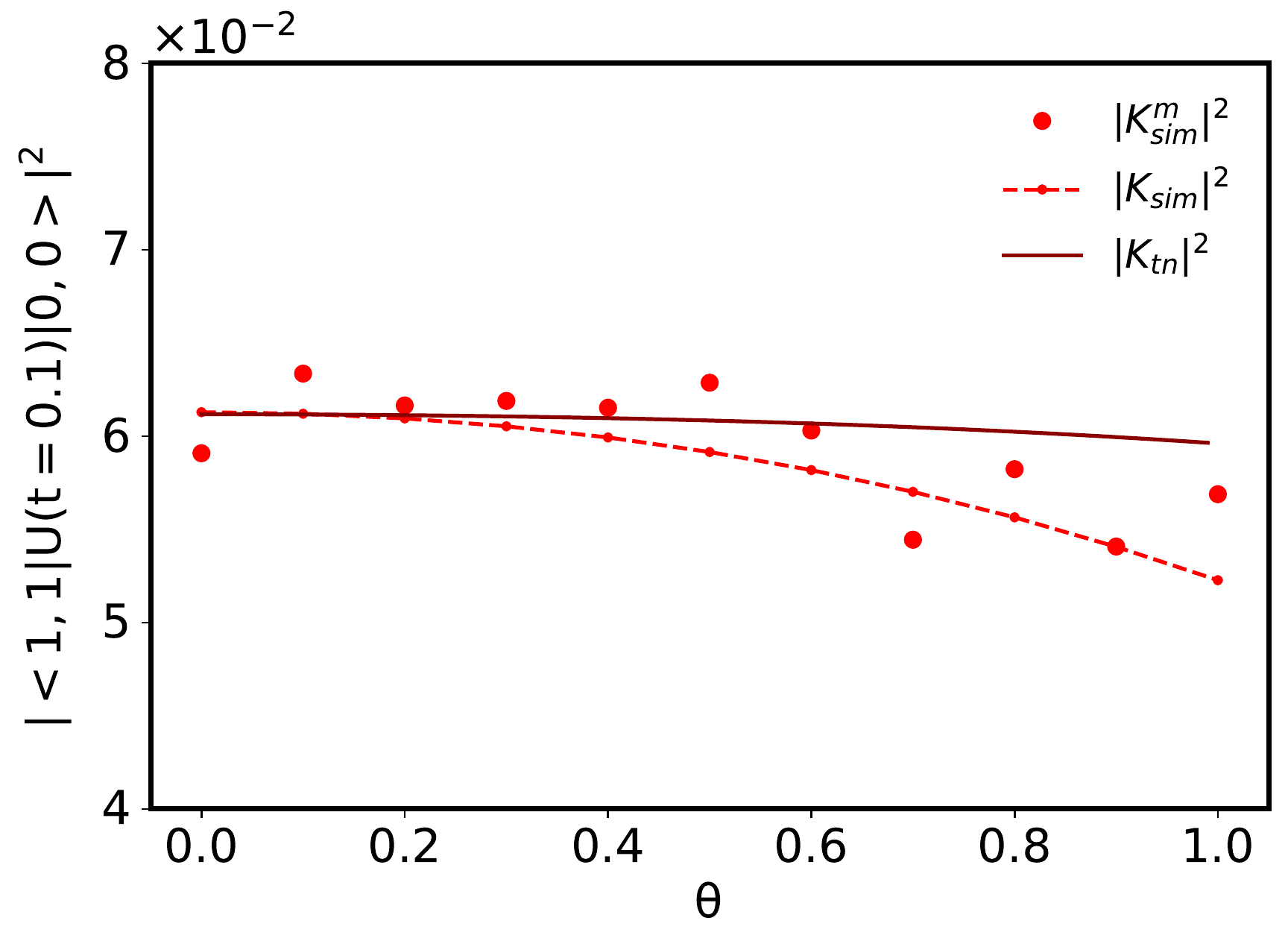}
       (c)
                \end{minipage}
        \begin{minipage}[c]{0.5\textwidth}\centering 
       \includegraphics[width=\textwidth]{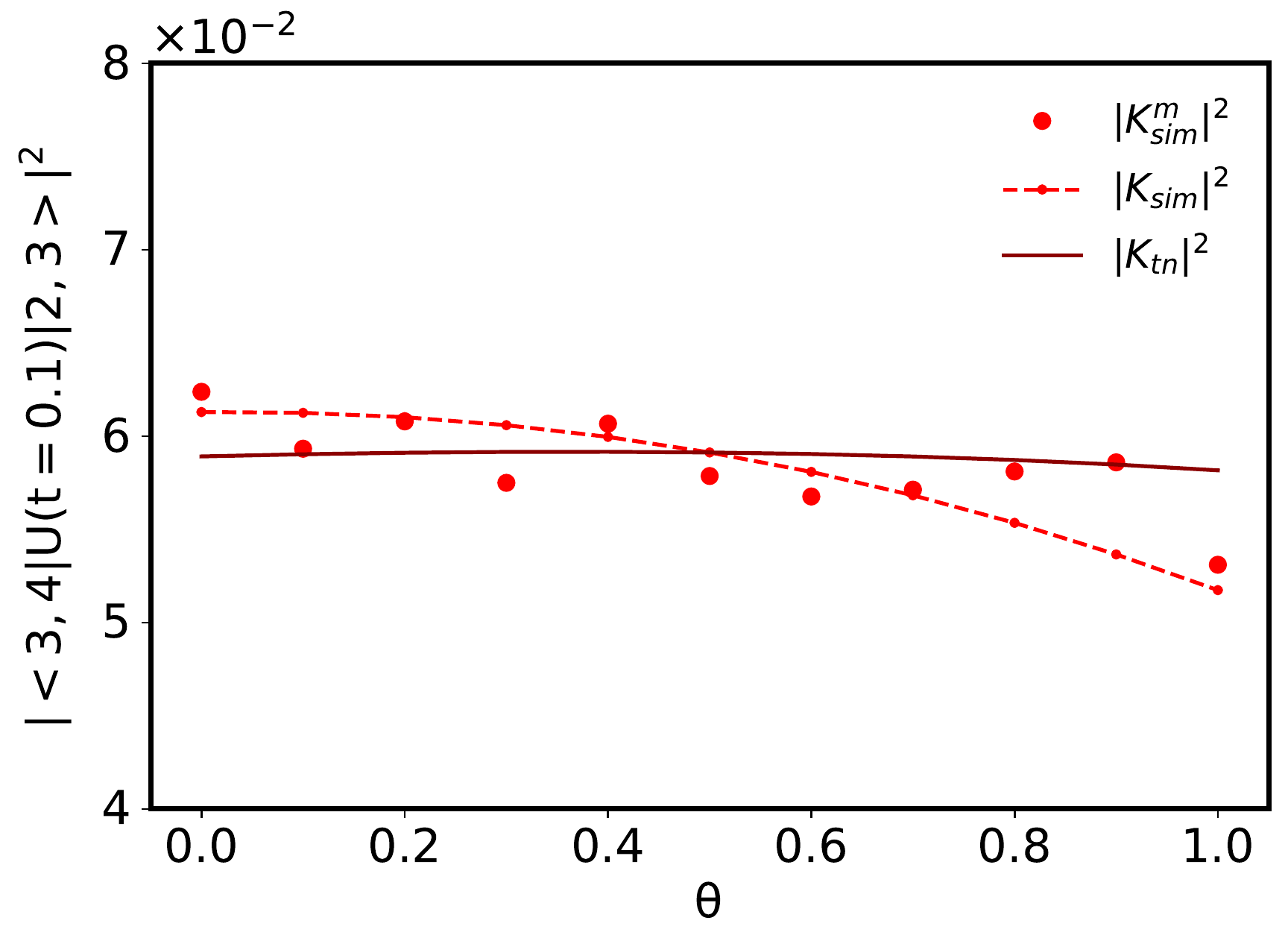}
       (d)
                \end{minipage}
         \caption{Variation of $|\bra{x',y'}U(t)\ket{x,y}|^2$ as a function of $\theta$ at time $t=0.1$.}
         \label{result2}
                \end{figure}

We take the discrete values of $x,\ y \in [-16,...,15]$ (each multiplied with $\sqrt{2\pi/N}$ with $N=32$) for which we require 10 qubits to express a position state $|x,y\rangle$ in terms of computational basis. Now using this Hamiltonian, we implement the time evolution of the quantum state of the two-dimensional harmonic oscillator, initially taken at the origin, $|0,0\rangle$ at $t=0$, and determine its probability distribution over the two-dimensional $32\times 32$ mesh at $t=0.2$ for two values of $\theta =0,1$ as shown in Fig. \ref{timefig}.
As it is challenging to disentangle the effect of $\theta$ on the simulated time evolution just from Fig. \ref{timefig}, first, by fixing the initial and final states, we calculate the overlapping between them directly and compare it with the result obtained from the measurement in the simulation which we present in Fig. \ref{result1} and \ref{result2}. 

Also, regarding the theoretical values of the time variation of the overlapping between different position states of the two-dimensional isotropic oscillator, which are included in Fig. \ref{result1} and \ref{result2}, we would like to point out a few subtle issues. First of all, the position and momentum operators of the Hamiltonian in Eq. (\ref{commhamiltonian}) for the isotropic case are unbounded Hermitian operators acting on the infinite-dimensional Hilbert space of square-integrable functions, $L^{2}(\mathbb{R}^{2},d x d y)$ whereas the implementation of the position and momenta operators in the quantum simulations are finite dimensional matrices. As a consequence, the commutations relations between these operators are not equal to some c-numbers. Furthermore, as shown in \cite{jagannathan, Singh:2017eav, Singh:2018qzk}, the finite-dimensional quantum mechanical systems are intrinsically different from the ordinary quantum mechanical systems with truncated Hilbert space. For example, the uniformly spaced eigenspectrum of the ordinary quantum harmonic oscillator does not hold for the finite-dimensional case. However, by considering the Hamiltonian as a difference operator (rather than differential operators), one can construct energy eigenfunctions of this discrete harmonic oscillator which are only defined on discrete equidistant points of space and admit the uniformly spaced eigenspectrum (see for example, \cite{Atakishiyev:1990, Atakishiyev:1998md, Lorente}). But such modification of the Hamiltonian is structurally different than the usual Hamiltonian of the harmonic oscillator, and the eigenfunctions also depend on additional parameters and the lattice configuration, and therefore it is not straightforward to adopt the eigenspectrum for different lattice structures. Because of such technicalities, we avoid such construction for finite-dimensional quantum systems.  Besides, the group theoretical formalism, which allowed us to map the noncommutative two-dimensional quantum mechanical system into the ordinary quantum mechanical system, presented in section \ref{grouptheoretical}, cannot be adopted for the finite-dimensional case. Furthermore, the exact propagator for the usual two-dimensional noncommutative isotropic oscillator (see appendix) significantly deviates from the simulation results that encompass the finite-dimensional system. Therefore, when comparing the results of the simulation with theory, we first calculate the overlapping between different position states by truncating the energy eigenvalues to a maximum value in the corresponding propagator for each case. Then we normalize it in such a way that at $t=0$, the overlapping between two same position states for the truncated case matches with the finite-dimensional case, which is given by $\langle f|i\rangle=\delta_{if}$. In Fig. \ref{result1} and \ref{result2}, we denote the theoretical values of the overlapping from normalized truncated propagator as $|K_{tn}|^{2}$. Also, after carrying out the simulation over a certain time, the overlapping between two different position states calculated directly and via measurements are denoted as $|K_{sim}|^{2}$ and $|K^{m}_{sim}|^{2}$, respectively. Here, the quantity $|K_{sim}|^2$ is obtained by calculating directly the inner product between the position eigenstates in Qiskit, whereas $|K_{sim}^m|^2$ is obtained by the counts of the considered position eigenstate appeared in the measurement of the final state with 8192 shots. From Fig. \ref{result1}, we can see that the presence of non-zero $\theta$ results in a slightly faster variation of the overlapping between different position states with time compared to the case of $\theta=0$ in the simulation, which is corroborated by the theoretical predictions, because the larger the $\theta$, the larger the frequency $\Omega$ becomes as stated in Eq. (\ref{isoharmpara}). 

\begin{figure}[H]
\centering
\hspace*{1.6cm}   
\includegraphics[width=10cm]{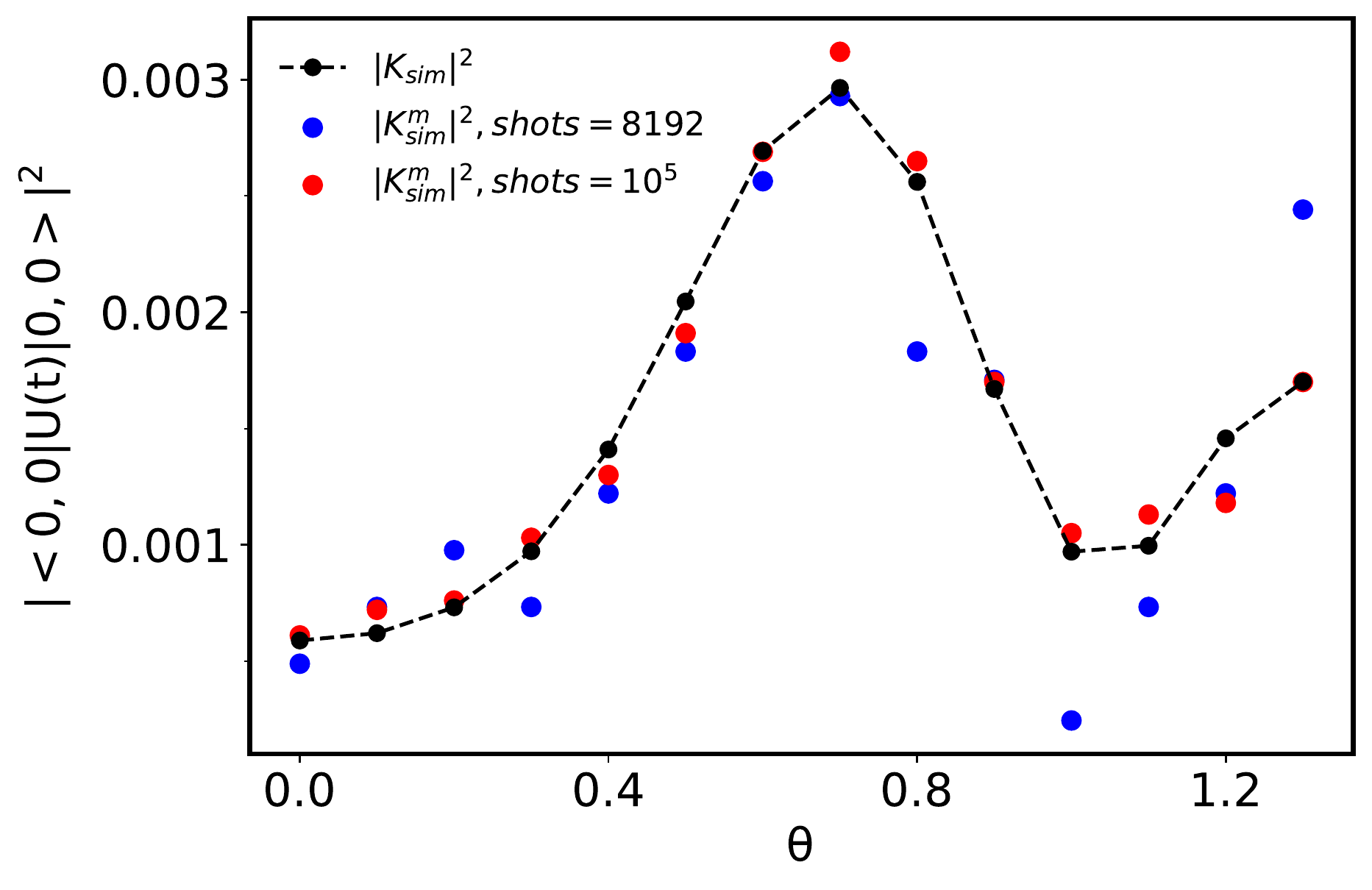}
\caption{Comparison of the overlapping $\langle 0,0|U(t)|0,0\rangle|^{2}$ among the directly calculated value $|K_{sim}|^2$ and the values obtained by measurement $|K_{sim}^m|^2$ with 8192 shots and $10^5$ shots, respectively, as a function of $\theta$.}
\label{simm}
\end{figure}

In addition, in Fig. \ref{result1} and \ref{result2}, we see fluctuations in the overlapping between position states determined by the measurement in the simulation. In this case, the number of counts  or how many times the quantum state (in general, the superposition of multiple states) after the simulation collapses onto the considered final position state determines by the overlapping between these position states. As the measurement process is inherently random, to mitigate the associated statistical fluctuation, we can increase the sample size by taking more shots, which is nothing but the number of times the quantum circuit is executed before obtaining the counts. It is evident from Fig. \ref{simm} that the deviation of the values obtained by measurement from the calculated values decreases as a higher number of shots is taken.

\begin{figure}[H]
\centering
\includegraphics[width=10cm]{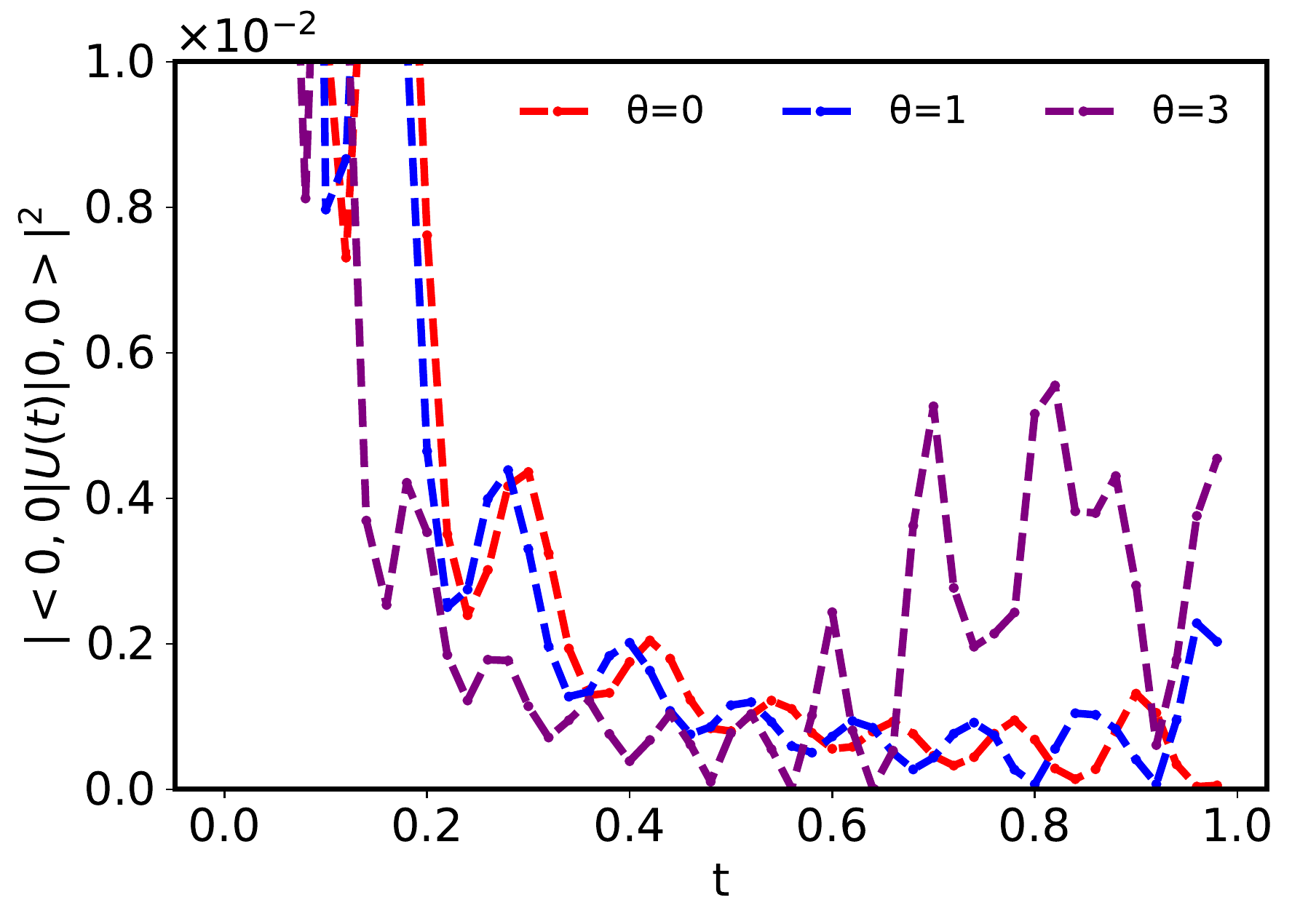}
\caption{Evaluation of the overlapping, $|\langle0,0|U(t)|0,0\rangle|^2$ as a function of time for three values of $\theta$, where we can see that at larger time $t\simgt 0.5$, the value of the overlapping for $\theta=3$ has started deviating from the cases for $\theta=0$ and 1, and it signals the degradation of the quality of the quantum simulation itself. Here, the values of the overlapping were directly obtained by calculating inner products between position eigenstates in Qiskit.}
\label{trotterlargetheta}
\end{figure}

Furthermore, let us elucidate on the Trotter error associated with the non-zero $\theta$. For the Hamiltonian $H=\sum_{i=1}^{\Gamma}H_{i}$ which can be decomposed into $\Gamma$ terms, the upper bound on the error coming from the approximation of the time evolution operator $U(t)=e^{-i H t}$ with first-order Trotter-Suzuki product formula $U^{(n)}(t)=\left(e^{-i H_{\Gamma}t/n}...e^{-i H_{1}t/n}\right)^{n}$, is given by
\begin{equation}
    \left\Vert U(t)-U^{(n)}(t)\right\Vert\leq \frac{t^{2}}{n}\left\Vert\sum_{i<j=1}^{\Gamma}[H_i,H_j]\right\Vert \;,
    \label{trottererr}
\end{equation}
here, $\Vert . \Vert$ is the matrix norm\footnote{Here, we consider Frobenius norm of a matrix $A$ which is given as $\Vert A \Vert =\sqrt{\mathrm{Tr}(A^{T}.A)}$.} and $n$ is the Trotter step. As we have simulated the time evolution of the two-dimensional isotropic harmonic oscillator with Hamiltonian Eq. (\ref{isoham}) (expressing it as $H_{X}=H_{1}+H_{2}$, $H_{Y}=H_{3}+H_{4}$ and $H_{XY}=H_{5}+H_{6}$) using this first-order product formula, the relevant terms associated with the Trotter error (for a detailed account, see Ref.~\cite{Childs_2021}) are,
\begin{align}
    \Vert [H_{1},H_{2}] \Vert &\sim \Vert [H_{3},H_{4}] \Vert \sim \Omega^2\sim m^{2}\omega^{4}\theta^{2}+\omega^{2},\nonumber\\
    \Vert [H_{1},H_{5}]\Vert &\sim \Vert [H_{3},H_{6}]\Vert \sim l/M\sim m^{2}\omega^{3}\theta^{4},\nonumber\\
    \Vert [H_{2},H_{6}]\Vert &\sim \Vert[H_{4},H_{5}]\Vert \sim l M\Omega^{2}\sim m^{2}\omega^{3}\theta^{2},\nonumber\\
   \Vert [H_{5},H_{6}]\Vert &\sim l^{2}\sim m^{2}\omega^{2}\theta^{4},\label{trotterdep}
\end{align}
where the rightmost terms are the approximations in the large $\theta$ limit. Thus we can see that the angular momentum term $H_{XY}=H_{5}+H_{6}$ in the Hamiltonian induced by the non-zero $\theta$ contributes to the Trotter error at most at the order $O(\theta^{4})$. Consequently, for larger time $t$, the increasing value of noncommutativity parameter $\theta$ will increase the Trotter error and degrade the quality of the quantum simulation itself. This behavior is demonstrated in Fig. \ref{trotterlargetheta}. As a consequence, for a reliable quantum simulation of a quantum mechanical system with spatial noncommutativity, one has to maintain the value of $\theta$ small so that roughly $c\,\theta/t^{2}\simlt 1$ where $c$ can be a dimensionless combination of other parameters of the quantum system itself.

\begin{figure}[H]
\centering
\includegraphics[width=8cm]{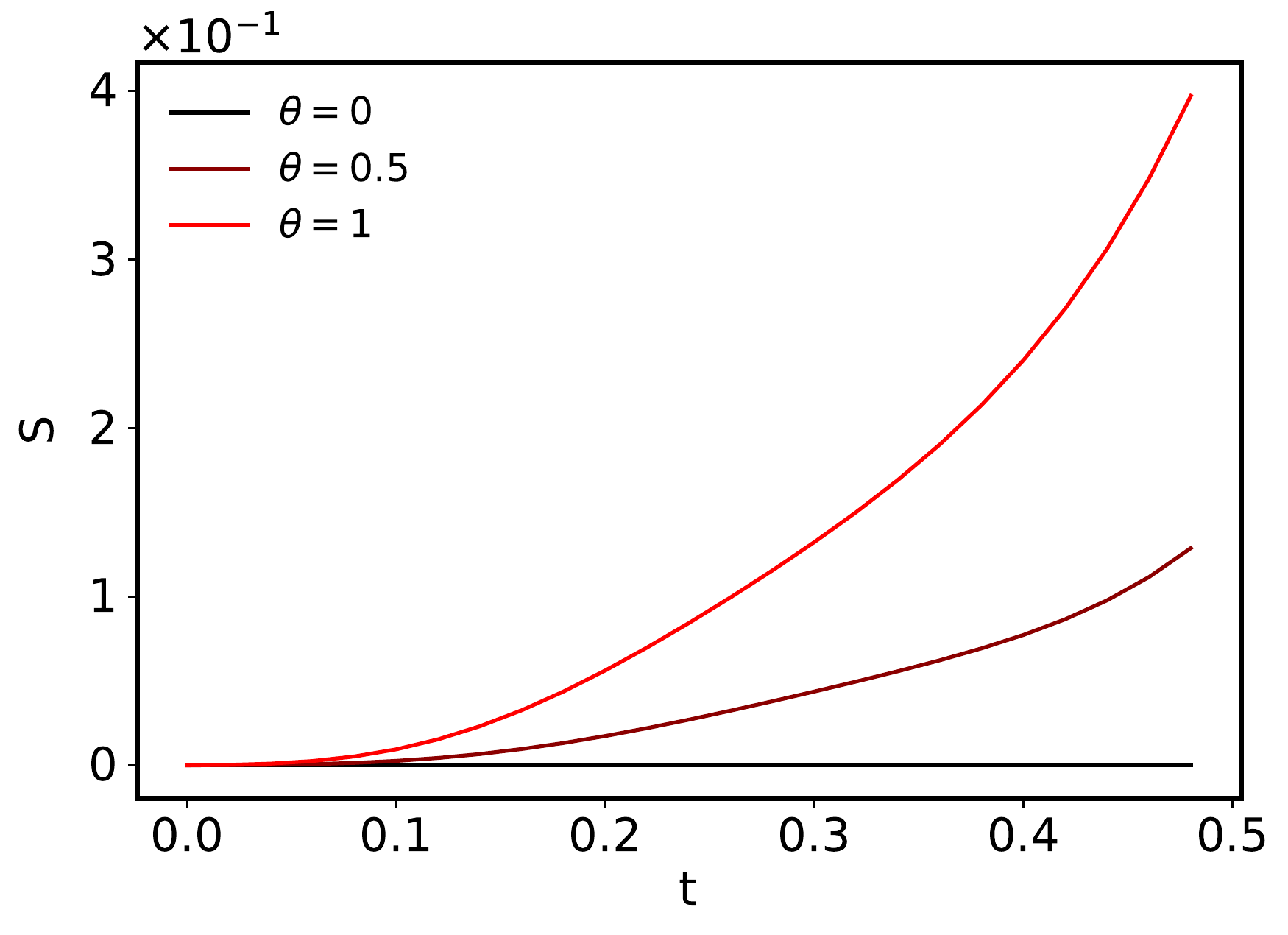}
\caption{Bipartite entanglement entropy S as a function of time t for noncommutativity parameter, $\theta=0, 0.5$ and $1$. We start with the initial state $\ket{x=0,y=0}$ and calculate the bipartite entanglement entropy between the qubits dedicated to x-axis and y-axis at each trotter step using the built-in functions of Qiskit \cite{Qiskit}.}
\label{entropy}
\end{figure}

Furthermore, the noncommutativity parameter $\theta$ introduces the term $H_{XY}$ in the Hamiltonian of the two-dimensional simple harmonic oscillator that results in the non-separability of the wavefunction in $x$ and $y$ dimensions. Therefore, the subspaces spanned by the qubits dedicated to the $x$ and $y$ dimensions experience a bipartite entanglement between them under the time evolution which is presented in Fig. \ref{entropy}. Here, the bipartite entanglement entropy is calculated as $S=\mathrm{Tr}[\rho_{X}\mathrm{log}\rho_{X}]=\mathrm{Tr}[\rho_{Y}\mathrm{log}\rho_{Y}]$ where $\rho_{X,Y}=\mathrm{Tr}_{Y,X}(|\psi\rangle \langle \psi|)$ (where, $|\psi\rangle$ is a general position state) are the reduced density matrices associated with the subspaces of qubits dedicated for $x$ and $y$ dimensions, respectively. We started from $|0,0\rangle$ position state and as expected, for $\theta=0$ we see no entanglement entropy and for $\theta\neq 0$ its gradual increase with time. Moreover, for larger values of $\theta$, larger entanglement entropy is generated with time.

\subsection{Noisy Simulation}\label{noisysimsec}
In this section, we analyze the effects of various types of noise, for example, the readout error, the depolarizing error, and the thermal relaxation error on the simulation following \cite{Georgopoulos:2021fyi}. We take $N=8$ to work with fewer qubits so that the circuit depth is small enough to discern the effects of different types of noise, as a higher number of qubits degrade the simulation quality quickly. We compare the overlapping $|\bra{0,0}U(t)\ket{0,0}|^2$ as a function of time for the ideal case and noisy simulation.
\begin{figure}[H]
 \begin{minipage}[c]{0.5\textwidth}\centering 
  \includegraphics[width=\textwidth]{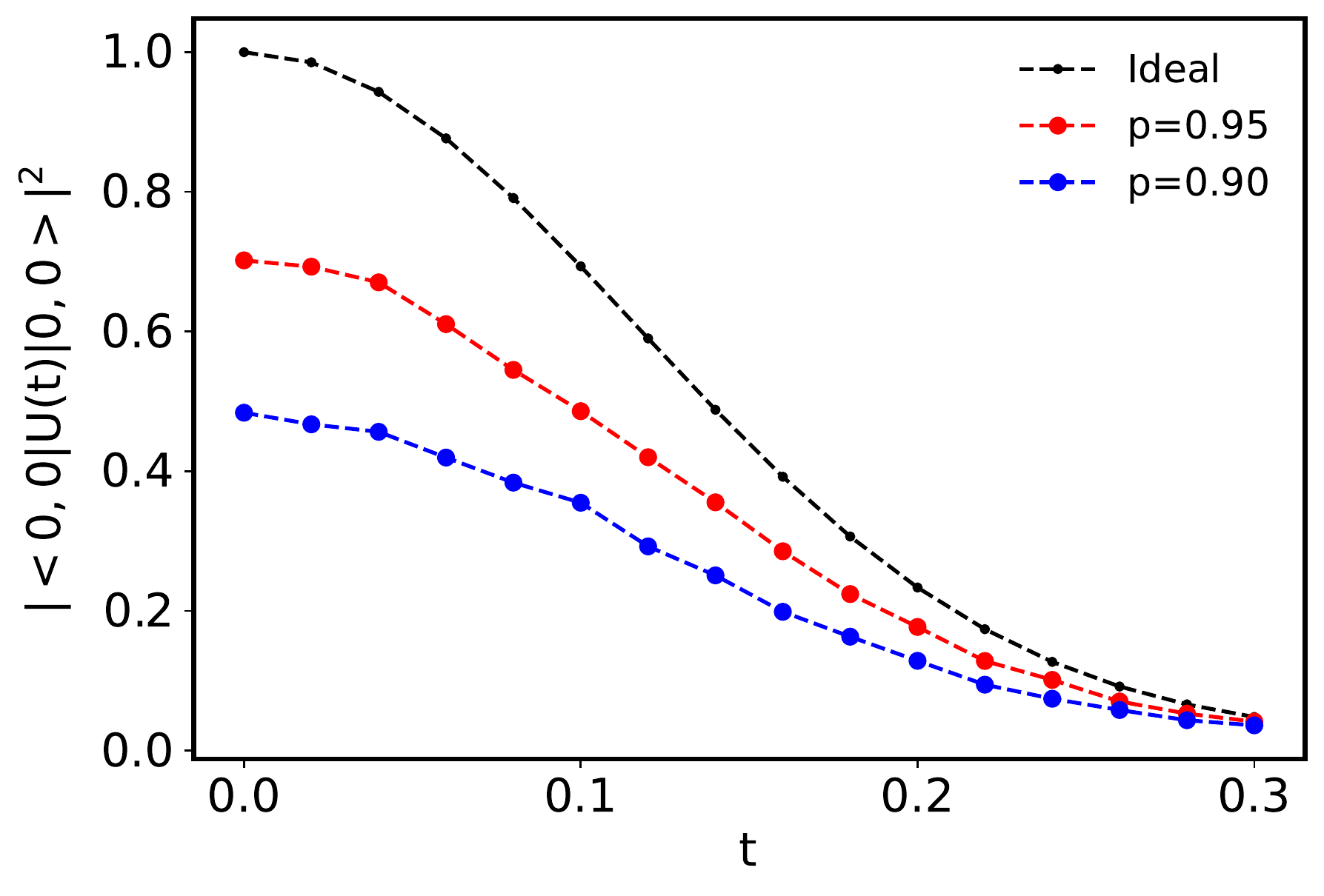}
  \begin{picture}(0,0)
      \put(-5,167){Readout Error}
       \end{picture}
  (a)
                \end{minipage}
         \begin{minipage}[c]{0.5\textwidth}\centering
         
       \includegraphics[width=\textwidth]{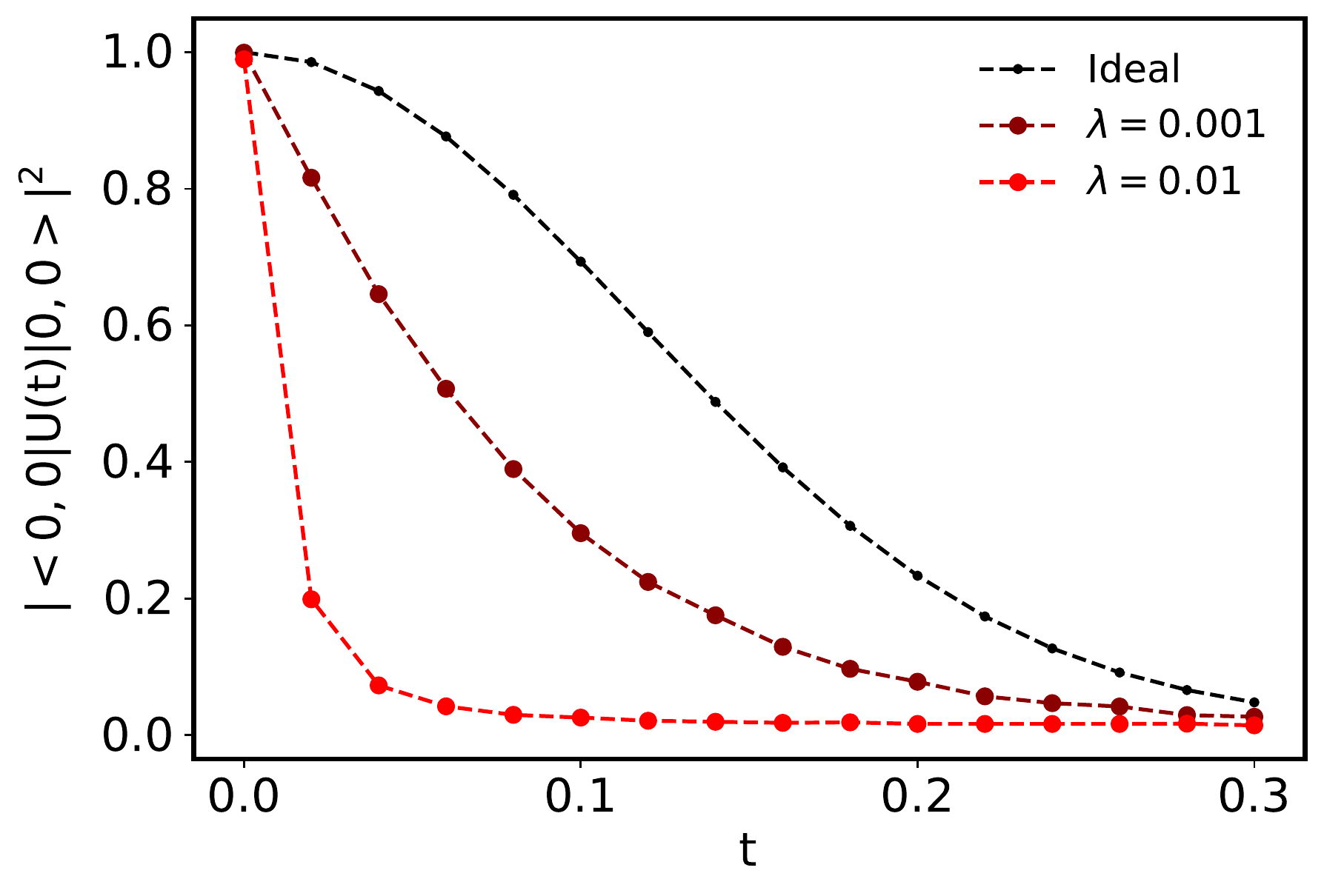}
       \begin{picture}(0,0)
      \put(-20,167){Depolarizing Error}
       \end{picture}
       (b)
                \end{minipage} \begin{minipage}[c]{0.5\textwidth}\centering 
       \includegraphics[width=\textwidth]{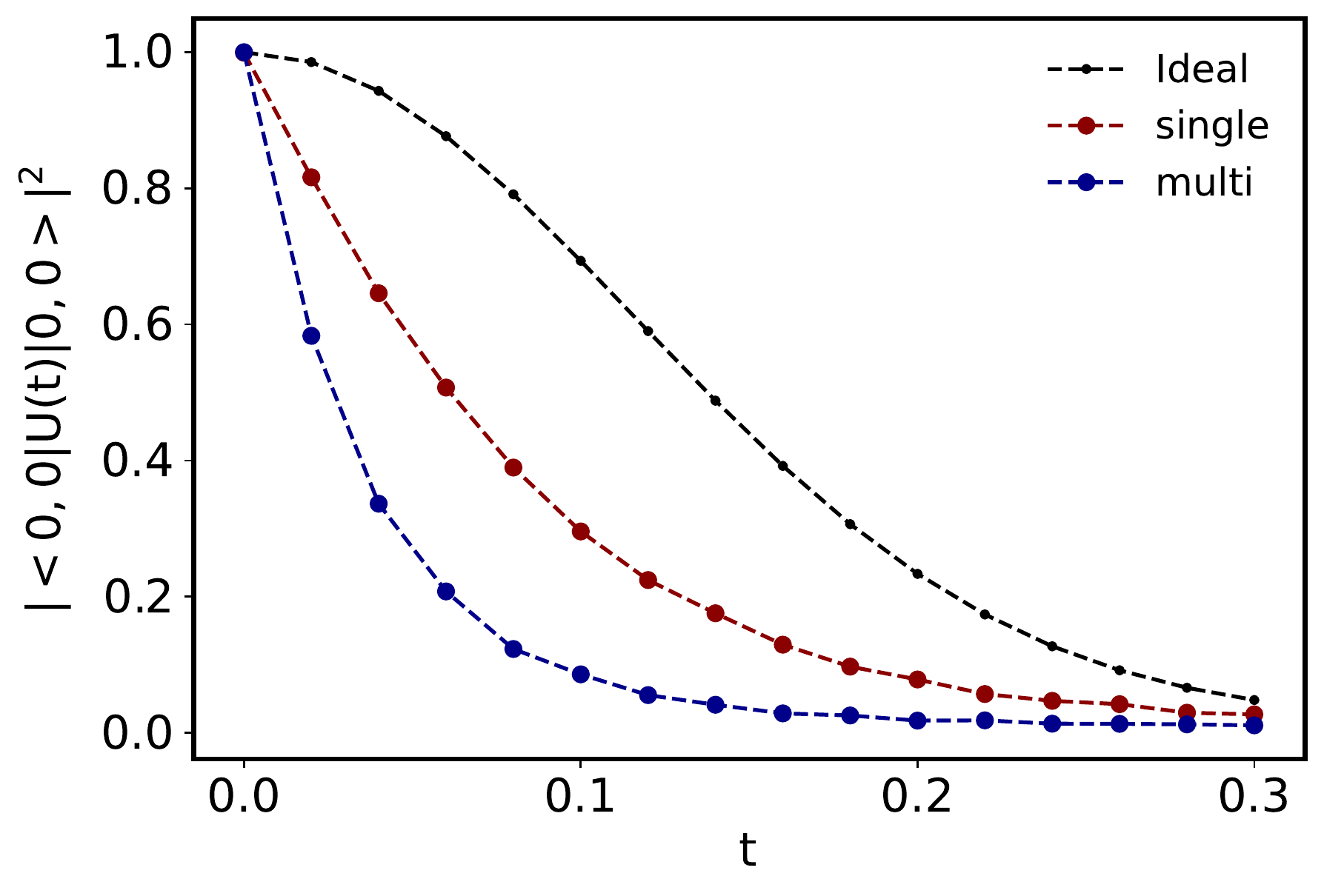}\begin{picture}(0,0)
      \put(-150,156){Depolarizing Error}
       \end{picture}
       (c)
                \end{minipage}
        \begin{minipage}[c]{0.5\textwidth}\centering 
       \includegraphics[width=\textwidth]{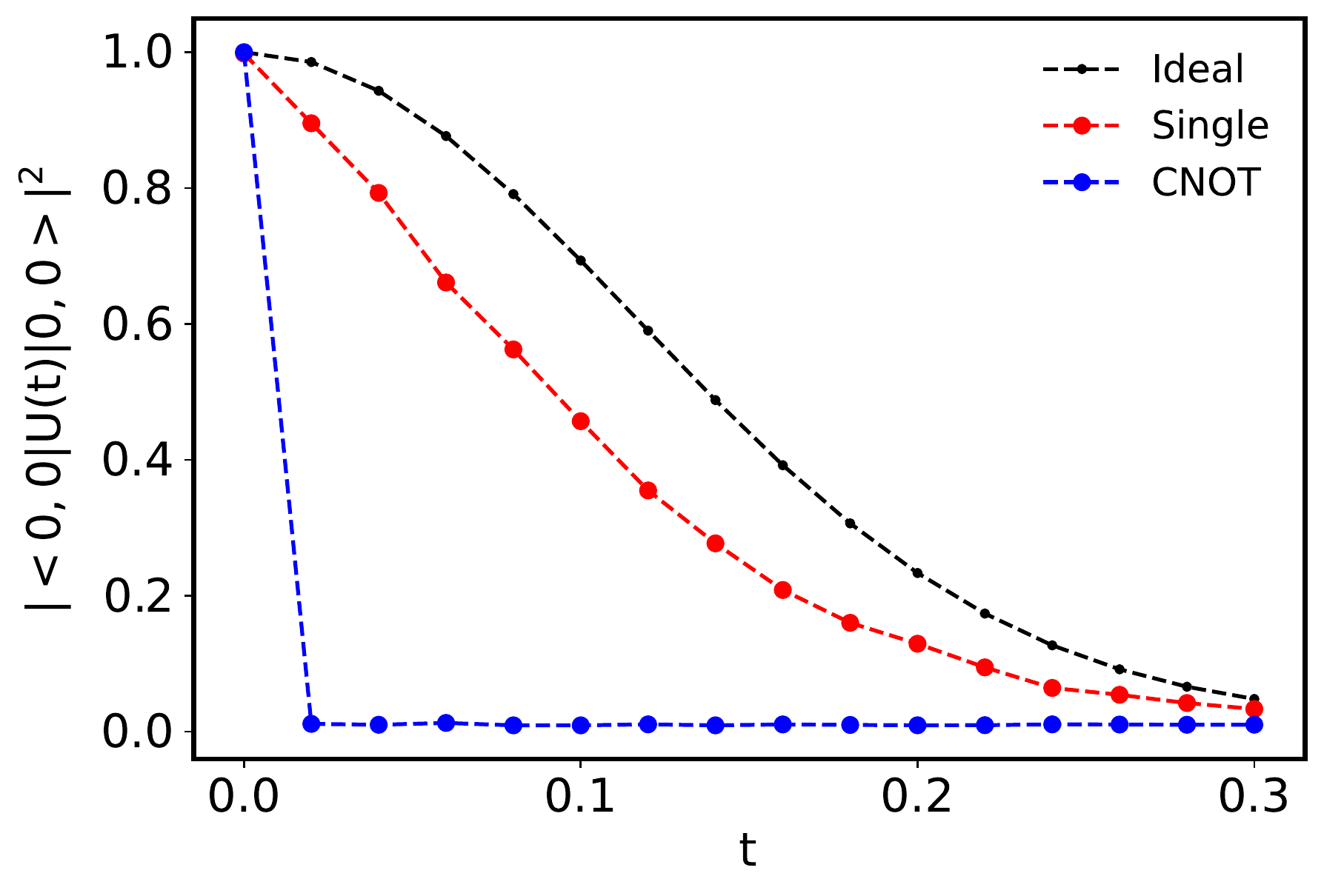}
       \begin{picture}(0,0)
      \put(-33,167){Thermal Relaxation Error}
       \end{picture}
       (d)
                \end{minipage}
\caption{Effect of various types of error on the simulation. Here, the overlapping $|\bra{0,0}U(t)\ket{0,0}|^2$ is plotted for the ideal case and the noisy simulation for $N=8$ qubits. Besides, $m=0.5$, $\omega=1$, $\theta=1$ and $\delta t=0.02$. (a) Readout error where p is the probability of correct output. (b) Depolarizing error introduced only on the single-qubit gates for the depolarization parameter $\lambda= 0.001$ and $0.01$. (c) Comparison of the depolarizing error of single-qubit gates and multi-qubit gates for $\lambda=0.001$. (d) Comparison of the thermal relaxation error of single-qubit gates and CNOT gates for $T_1=50\,\mu s$, $T_2=100 \,\mu s$, $T_{sq}=30\,ns$, $T_{CNOT}=300\,ns$.}
\label{error}
                \end{figure}
Firstly, we add a readout error to each qubit with the probability of correct output being $p$. The effect of this error is to reduce the value of the overlapping without altering the shape of the overlapping vs. time curve since measuring each qubit returns the incorrect value with a probability $1-p$, thus reducing the amplitude of the correct state without affecting the dynamics, as seen in Fig. \ref{error} (a).

Next, we study the effect of the depolarizing error. Depolarization occurs due to bit-flip and phase-flip errors, leading to a completely mixed state. We compare the effect of depolarization on the single-qubit gates for the values $\lambda=0.001$ and $0.01$ where even with a slight deviation from the ideal case, $\lambda=0$, the simulation deteriorates drastically, as shown in Fig. \ref{error} (b). Besides, for $\lambda=0.001$, we compare the depolarization error of single-qubit and multi-qubit gates, and we find that the error is more significant for the multi-qubit gates compared to the single-qubit gates, as seen in Fig. \ref{error}(c).

Lastly, we analyze the effect of thermal relaxation error on our simulation. Thermal relaxation error is characterized by two constants, $T_1$ and $T_2$, where $T_1$ refers to the time a qubit decays from an excited state to the ground state, and $T_2$ is dephasing time characterizing the loss of coherence of a quantum state. These constants limit the number of operations performed on a qubit. We take $T_1$ and $T_2$ from a Gaussian distribution with mean $50\,\mu s$ and $100\,\mu s$, respectively with a standard deviation of $10\,\mu s$. We take the single-qubit execution time to be $T_{sq}=30\,ns$ and CNOT gate execution time $T_{CNOT}$ to be $300\,ns$. These values are approximated from the values obtained from current backends offered by IBM. Here, we only compare the thermal relaxation error of the single-qubit gates and the CNOT gates. Besides, the multi-qubit gates can be transpiled into single-qubit and CNOT gates. As seen in Fig. \ref{error} (d), the thermal relaxation error on the CNOT gates is much more prominent than the single-qubit case due to many CNOT gates and a ten times larger execution time. Therefore, we conclude that the circuit depth corresponding to our case of $N=32$ qubits is too large to execute a noisy simulation in a NISQ device.

\section{Conclusion and Outlook}\label{conclusion}
In this work, we carry out a noiseless quantum simulation of the two-dimensional isotropic quantum harmonic oscillator with spatial noncommutativity, which is a representative quantum system of NCQM in a quantum computer setup provided by Qiskit. First, we use the group theoretical method to map the Hamiltonian of the two-dimensional NCQM into the ordinary quantum mechanical Hamiltonian. Afterward, we discretize the two-dimensional space into $32\times 32$ mesh where $x,y\in [-16,...,15]$ and set up the corresponding computational basis with $n=2\,\mathrm{log}_{2}32=10$ qubits for the $2 D$ discrete points. Next, we construct the position operator as a diagonal matrix for each dimension and the momentum operator as the centered discrete Fourier transform of the position matrix, respectively, both acting on the computational basis. Now equipped with the Hamiltonian, we assemble the quantum gate that implements the time evolution using the Trotter-Suzuki formula. One can easily extend this framework for higher $d$-dimensional quantum systems that have translational symmetry with a set of $p=d\,\mathrm{log}_{2}N$ qubits since the position and momentum operators are identical for each dimension.

As the noncommutativity parameter $\theta$ increases the overall frequency of the harmonic oscillator, we see the discerning shift in the time variation of the overlapping between different position eigenstates in our simulation, presented in Fig. \ref{result1} for a smaller value of time compared to the case of $\theta=0$. In addition, for a fixed time, the variation of the overlapping between different position eigenstates with respect to $\theta$ is also observed in our simulation, as shown in Fig. \ref{result2}. However, the discretization of space required to implement our quantum simulation leads to the incompatibility of position and momentum operators acting on the finite-dimensional Hilbert space with the usual unbounded self-adjoint position and momentum operators of the ordinary quantum mechanical system working on the infinite-dimensional Hilbert space  or even its truncated version. As the group theoretical formalism to map the NCQM to ordinary quantum mechanics is not available yet for a finite-dimensional quantum mechanical system, we use the truncated propagator to calculate the overlapping between different position states and, as a result, acquire some error which is seen in both Fig. \ref{result1} and \ref{result2} as the small deviation between theoretical prediction and the values obtained from the simulation. The effects of discretization can be minimized by smaller lattice spacing, but for the computational basis, this translates into taking a large number of qubits.

Moreover, the main obstacle to executing the circuit in real devices is its higher circuit depth due to a large number of trotter steps. Even noisy simulations using the noise models produce highly erroneous results, from which we conclude that the simulation is not suitable for the current NISQ devices. Hence, we opted for the noiseless simulation in the qasm simulator of Qiskit. Likewise, we consider 8192 shots to produce the probability distribution of position states with time in our simulation that can be extended up to $10^{5}$ in Qiskit currently, but the higher number of shots, though reducing the statistical error, increases the execution time. Moreover, for $N=32$ qubits representing a single two-dimensional oscillator on the noncommutative space, the algorithm we followed requires $O(10^5)$ gates for a single Trotter step. So if we have $n$ time steps, the number of gates scale as $O(10^5 n)$. Also, if we extend our scenario of a single oscillator to  $L$ number of oscillators, the resource would scale as $O(10^5 n L)$ if we neglect the coupling between the oscillators for simplicity.

Besides, the simulation is implemented using the first-order Trotter-Suzuki product formula for which the presence of spatial noncommutativity parameter induces additional large errors when its value is taken larger as seen from the estimates given in Eq. (\ref{trotterdep}) and Fig. \ref{trotterlargetheta}. Hence, the Hamiltonian simulation via this method is limited to a range of optimal values of time and parameters. Nevertheless, there are algorithms for Hamiltonian simulation based on continuous and fractional queries \cite{berry2014exponential}, Taylor series \cite{berry2015simulating}, and quantum walk \cite{berry2009black}. Quantum walk-based algorithms have query complexities that are better scaled than product formulas with the Hamiltonian's time of evolution and sparsity, whereas the fractional query model simulation has better scaling with the allowed error in the Hamiltonian. Hamiltonian simulation combining these two approaches to obtain optimal dependence on sparsity, error, and evolution time is given in \cite{berry2015hamiltonian}. It will be interesting to carry out the Hamiltonian simulation of quantum systems with additional noncommutativity parameters using the above-mentioned algorithms and find out the impact of these parameters on the simulation.

\section*{Acknowledgement}
TAC would like to thank the High Energy Theory group of the Department of Physics and Astronomy at the University of Kansas for the hospitality and support. The work of S.N is supported by the United Arab Emirates University under UPAR Grant No. 12S093. We acknowledge the use of IBM Quantum services for this work. The views expressed are those of the authors and do not reflect the official policy or position of IBM or the IBM Quantum team.

\appendix
\section{The isotropic two-dimensional harmonic oscillator with spatial noncommutativity}
In the isotropic limit, the Hamiltonian of the two-dimensional harmonic oscillator containing noncommutativity parameter $\theta$, given in Eq. (\ref{commhamiltonian}) has $M_{1}=M_{2}=M$, $\Omega_{1}=\Omega_{2}=\Omega$ and $l_{1}=l_{2}=l$, and it is separable in the polar coordinates ($x=r\cos\phi,\,y=r\sin\phi$) because of the $\theta$ induced angular momentum operator, $\hat{L}_{z}=\hat{x}\hat{p}_{y}-\hat{y}\hat{p}_{x}$. The energy eigenfunctions can be readily found which are given by,
\begin{equation}
    \psi_{n_r,m_l}(r,\phi)=\sqrt{\frac{M\Omega}{\pi\hbar}}\sqrt{\frac{n_r!}{(n_r+|m_l|)!}}\left(\frac{M\Omega}{\hbar}\right)^{\frac{|m_l|}{2}}r^{|m_l|}e^{-\frac{M\Omega}{2\hbar}r^2}L_{n_r}^{|m_l|}\left(\frac{M\Omega}{\hbar}r^2\right)
    e^{im_l\phi},
    \label{wf}
\end{equation}
with the energy eigenvalues
\begin{equation}
    E_{n_r,m_l}=\hbar\Omega(2n_r+|m_l|+1)+m_l\,l\hbar ,
\end{equation}
where $n_r=0,1,2,...$  and $m_l=0,\pm1,\pm2,...$ and the total energy eigenvalue is $n_{tot}=2 n_{r}+ |m_{l}|$.

Now we identify the position eigenstate $|x,y\rangle=|r\cos\phi,r\sin\phi\rangle\equiv|r,\phi\rangle$. The exact propagator $\langle r',\phi'|e^{-i H t/\hbar}|r,\phi\rangle$ denoted by $K(r',\phi',r,\phi,t)\equiv K(x',y',x,y,t)$ is also easily found to be, 
\begin{equation}
    K(r',\phi',r,\phi,t)=\frac{M\Omega}{2\pi i\hbar sin(\Omega t)} exp\left(-\frac{M \Omega}{2i\hbar}\frac{cos(\Omega t)}{sin(\Omega t)}(r'^2+r^2)\right) exp\left(-i\frac{M\Omega}{\hbar}\frac{1}{sin(\Omega t)}rr'cos(\phi'-\phi+lt)\right).
\end{equation}

As the exact propagator is inadequate to compare with the values obtained in the quantum simulation, we consider the truncated propagator up to a maximum value of total energy eigenvalue, $n^{\mathrm{max}}_{tot}$.
\begin{equation}
   K_{t}(x',y',x,y,t)\equiv K_{t}(r',\phi',r,\phi,t)=\sum_{n_r=0}^{n_r^{max}}\sum_{m_l=-m_l^{max}}^{m_l^{max}}e^{-iE_{n_r,m_l}t/\hbar}\psi^{*}_{n_r,m_l}(r',\phi')\psi_{n_r,m_l}(r,\phi).
\end{equation}

\bibliographystyle{style}
\bibliography{references}
\end{document}